\documentclass[a4paper,10pt]{article}
\pdfoutput=1
\usepackage{fullpage}
\usepackage{amsmath}
\usepackage{amssymb}
\usepackage{amsfonts}
\usepackage{graphicx}
\usepackage{caption}
\usepackage{subcaption}
\usepackage{float}
\usepackage[usenames,dvipsnames]{xcolor}
\usepackage{tikz}
\usepackage{pstricks}
\usepackage{authblk}
\usepackage{setspace}
\usepackage[pagebackref=true]{hyperref}
\hypersetup{
    colorlinks=true,
    linkcolor=blue,
    filecolor=Violet,     
    urlcolor=RubineRed,
    citecolor=red
}

\usepackage{array}
\usepackage[leftcaption]{sidecap}
\usepackage{tabu}
\usepackage{boldline,multirow}

\newcolumntype{C}[1]{>{\centering}m{#1}}

\begin{document}





\date{}

\author{Pritam Banerjee \thanks{\href{mailto:bpritam@iitk.ac.in}{bpritam@iitk.ac.in}}}

\author{Suvankar Paul \thanks{\href{mailto:svnkr@iitk.ac.in}{svnkr@iitk.ac.in}}}

\author{Rajibul Shaikh \thanks{\href{mailto:rshaikh@iitk.ac.in}{rshaikh@iitk.ac.in}}}

\author{Tapobrata Sarkar \thanks{\href{mailto:tapo@iitk.ac.in}{tapo@iitk.ac.in}}}

\affil{Department of Physics,\\  Indian Institute of Technology,  \\ Kanpur 208016, India}

\title{ Tidal effects away from the equatorial plane in Kerr backgrounds }


\maketitle

\begin{abstract}

We study tidal effects on self-gravitating Newtonian stars rotating around a Kerr black hole in stable circular orbits 
away from the equatorial plane. Such cases are exemplified by a non-vanishing Carter's constant. Here, we calculate the tidal 
disruption limit (Roche limit) of the star numerically, in Fermi normal coordinates. The Roche limit is found to depend 
strongly on the choice of the orbit, and differs significantly from the equatorial plane result as one approaches 
nearly polar orbits. As expected, this difference is large when the star is close to the black hole (near to the innermost 
stable circular orbit) and becomes smaller when the star is far from it. We also discuss the 
dependence of the Roche limit on the equation of state of the star, taking two specific parameter values as examples. 

\end{abstract}
\section{Introduction}

Black holes (BHs) are known to be the most compact objects of our universe. The gravitational field around their vicinity is so large that 
they can tidally disrupt compact objects such as neutron stars, white dwarfs etc. 
Tidal disruptions of stars produce some of the most fascinating astrophysical phenomena related to BHs. In fact, 
stellar objects that are tidally disrupted by black holes form the principal ingredients of accretion disks around them. 
This process may also give rise to a plethora of interesting phenomena with observational signatures, 
such as the creation of high energy gamma-ray bursts, formation of ultraviolet flare of a characteristic light-curve 
(see e.g \cite{uv-flare1},\cite{uv-flare2}) etc. Excellent reviews on the formation of gamma-ray bursts from BH-white 
dwarf mergers and BH-neutron star mergers can be found in \cite{GRB-BH-WD} and \cite{GRB-BH-NS}, respectively
(see also \cite{Rezzolla}). 

Recent discoveries of gravitational waves by binary BH mergers and binary neutron star mergers have also put the 
issue of tidal disruptions in the frontier of the study of black hole physics. 
If one can detect gravitational waves from tidal disruption events of neutron stars by BHs, 
it might help us to understand different features of BHs, as well as to constrain the neutron star equation of state. 
As a result, theoretical studies of tidal disruption of stars by BHs continue to be important in their own right.

The literature on the subject of tidal disruptions of stellar objects in the Newtonian and post-Newtonian 
approximation of gravity is, by now, vast (see e.g \cite{uv-flare1},\cite{tidal-literature1},\cite{Marck-tidal-literature2},\cite{tidal-literature3},
\cite{Ishii-Numerical}, etc.). However, to study tidal effects near a BH, we will need to take into account the 
full general relativistic effects on the stars. In this context too, there are several well known studies that exist in the literature 
(see \cite{Shibata-tidal},\cite{Marck-tidal},\cite{Fishbone},\cite{Ishii-Kerr}, etc.). 
In \cite{Marck-tidal} and \cite{Fishbone}, the tidal potential was calculated by using the geodesic deviation equation. 
On the other hand, Ishii, Shibata and Mino \cite{Ishii-Kerr}, evaluated this potential 
directly by using the tidal metric or the $ tt $-component of the Kerr metric 
expressed in terms of Fermi Normal Coordinates (FNCs) following Manasse and Misner \cite{Manasse-Misner}. 

Recall that a coordinate system describing a locally inertial frame which can be parallel-transported along the entire 
time-like geodesic of the star's motion is known as a Fermi Normal Coordinate system. In this paper, we will 
mostly use the methodology of Ishii, Shibata and Mino \cite{Ishii-Kerr}. We consider a compact star with a polytropic equation of 
state rotating around a Kerr BH in stable circular orbits. The orbital radius of the star around the BH is $ r $, 
and the average radius of the star (only due to its self gravity) 
is taken to be $ R_0 $. In the tidal approximation, we assume that 
$ R_0 < r $, and therefore the tidal potential on the star by the BH is expanded up to, say, fourth order in $ R_0/r $. It is known
that the third and fourth order terms play significant roles when $ R_0/r > 0.1$ \cite{Ishii-Kerr} and will be important for us. 
We will also incorporate the gravito-magnetic effects in addition to the tidal potential to obtain the tidal 
disruption limit or Roche limit of the star. In \cite{Ishii-Kerr}, the analysis was confined to stable circular motion of the star on the 
equatorial plane. In this paper, we generalize this by including 
the circular motion of the star for non-equatorial planes too, with the equatorial plane results arising as a limiting case. 
The assumption that we make here is that the star itself does not deform the Kerr background, i.e 
back-reaction effects are neglected. 

The motivation for this study is two fold. Apart from being theoretically interesting, note that 
stellar orbits away from the equatorial plane are more realistic compared to
the ones confined to that plane, since the Kerr BH possesses cylindrical symmetry. Indeed, this might have
significant relevance in futuristic analyses of gravitational waves arising out of mergers of black holes and compact stars. 
As we will see in sequel, our results indicate that there might be important differences on the nature of tidal disruptions 
of celestial objects off the equatorial plane, compared to the ones on it. Secondly, it is of interest to study the deformation of
stars due to gravity, in planes away from the equator. As we will see, there is a non-trivial effect that arises here in the context
of the Kerr BH, namely that (up to the order of approximation that we consider) 
the deformation of the star is not towards the black hole, but along a direction that varies with
the angular inclination of the orbit. As we will show, this can be explained by taking into account the net 
gravitational force on the stellar object. 

We mention at this point that the technical difficulty in the study of tidal forces in non-equatorial circular orbits in the Kerr 
BH backgrounds arises due to the presence of a non-zero Carter's constant. However, such orbits have been 
discussed in many works (see, e.g \cite{Wilkins}, \cite{Hughes1},\cite{Hughes2},\cite{Ryan}). Studying tidal effects in such 
orbits in Fermi normal coordinates involves a consistant numerical analysis, taking into account the various relevant 
parameters that appear. This is the study that we undertake in this paper.

The paper is arranged in the following order. In \hyperref[sec-2]{Section-2}, we review the characteristics of circular 
trajectories of massive objects (treated as effective point particles) 
in the background of Kerr BHs. This section is further divided into two sub-sections. 
\hyperref[sec-2.1]{Section-2.1} gives a brief description of stable circular orbits on the equatorial plane, 
whereas in \hyperref[sec-2.2]{Section-2.2}, we give a description of non-equatorial circular orbits, the 
inclination angle of the orbit with respect to the equatorial plane, relations between different constants of motion,
etc. In \hyperref[sec-3]{Section-3}, which is the main part of this paper, we discuss tidal effects in non-equatorial planes. 
This section is again divided into three subsections. \hyperref[sec-3.1]{Section-3.1} deals with the formulation of the problem. 
The hydrodynamic equation of the fluid star, expansion of the tidal potential up to fourth order in Fermi Normal Coordinates, 
and the two coupled equations that constitute the mathematical statement of the problem are described in this subsection. 
This is followed by \hyperref[sec-3.2]{Section-3.2}, where we convert the relevant equations into dimensionless ones, 
describe the numerical routine used in our analysis, define the Roche limit of the star etc. 
After that we present the main results of our numerical computations, and discuss them in \hyperref[sec-3.3]{Section-3.3}. 
We conclude our study with a summary and discussion in \hyperref[sec-4]{Section-4}.


\section{Circular Trajectories of massive objects in Kerr black hole}
\label{sec-2}

The metric of the Kerr space-time in Boyer-Lindquist coordinates is well known, and is given by
\begin{equation}
	ds^2 = - \left( 1-\frac{2 M r}{\Sigma} \right) dt^2 -\frac{4 M r a \sin^2 \theta}{\Sigma} dt d\phi + \frac{\Sigma}{\Delta} dr^2 + \Sigma d\theta^2 + \left( r^2 + a^2 + \frac{2 M r a^2 \sin^2 \theta}{\Sigma} \right) \sin^2 \theta d\phi^2 \label{eq.kerr-metric}
\end{equation}
where $ \Sigma = r^2 + a^2 \cos^2 \theta $ and $ \Delta = r^2 + a^2 -2 M r $.
Integration of geodesic equations for this metric gives \cite{Bardeen}
\begin{eqnarray}
	\Sigma \frac{dt}{d\tau} &=& E \left[ \frac{\left(r^2 + a^2 \right)^2}{\Delta} -a^2 \sin^2 \theta \right] +
	a L \left(1 - \frac{r^2 + a^2}{\Delta}\right) \nonumber\\
	\Sigma^2 \left(\frac{dr}{d\tau} \right)^2 &=& \left[E\left(r^2 + a^2\right) - a L\right]^2 - \Delta \left[r^2 + \left(L - a E\right)^2 + Q\right] = R(r) 
	\nonumber\\
	\Sigma^2 \left(\frac{d\theta}{d\tau} \right)^2 &=& Q - L^2 \cot^2 \theta - a^2 \left(1 - E^2\right) \cos^2 \theta = \Theta \nonumber \\
	\Sigma \frac{d\phi}{d\tau} &=& \frac{L}{\sin^2 \theta} + a E \left(\frac{r^2 + a^2}{\Delta} - 1\right) - \frac{a^2 L}{\Delta}
\label{eq.theta-dot}
\end{eqnarray}
Here, the quantities $ E $ and $ L $ represent the energy and the $ z $-component of angular momentum 
(per unit rest-mass) respectively, and $ Q $ represents the Carter constant (per unit rest-mass squared). 
These remain conserved along a specific geodesic. Let us now review the properties of stable circular orbits on and 
off the equatorial plane.


\subsection{Stable circular orbits : equatorial plane}
\label{sec-2.1}

If an object starts moving on the equatorial plane of the Kerr BH and remains on the same plane throughout, 
its orbits will always have $ \theta = \frac{\pi}{2} $ and $ \dot{\theta}=0 $. From the third relation of Eq.(\ref{eq.theta-dot}), 
this gives $ Q=0 $. So the value of Carter's constant for the equatorial orbits is zero, which is a necessary condition 
for the orbits to be confined on the equatorial plane. Again, in case of circular orbits, the velocity as well as the 
acceleration along the radial direction must vanish. This implies, for circular orbits, 
$ \frac{dr}{d\tau} = R(r) = 0 $ and $ \frac{d^2 r}{d\tau^2} = \frac{dR(r)}{dr} = 0 $. The solution of this set of equations 
defines $ E(r) $ and $ L(r) $ for both stable and unstable equatorial circular orbits. In case 
of stable orbits $ R''(r) < 0 $ and in case of unstable orbits $ R''(r) > 0 $. For the stable case, one obtains \cite{Bardeen},\cite{Hughes1}

\begin{equation}
	E^{pro}(r) = \frac{r^{3/2} - 2 M r^{1/2} + a \sqrt{M}}{r^{3/4}(r^{3/2} - 3 M r^{1/2} + 2 a \sqrt{M})^{1/2}}, ~~~ L^{pro}(r) = \frac{\sqrt{M}(r^2 + a^2 - 2 a \sqrt{M r})}{r^{3/4}(r^{3/2} - 3 M r^{1/2} + 2 a \sqrt{M})^{1/2}} \label{eq.E-L-prograde}
\end{equation}
\begin{equation}
	E^{ret}(r) = \frac{r^{3/2} - 2 M r^{1/2} - a \sqrt{M}}{r^{3/4}(r^{3/2} - 3 M r^{1/2} - 2 a \sqrt{M})^{1/2}}, ~~~ L^{ret}(r) = - \frac{\sqrt{M}(r^2 + a^2 + 2 a \sqrt{M r})}{r^{3/4}(r^{3/2} - 3 M r^{1/2} - 2 a \sqrt{M})^{1/2}} \label{eq.E-L-retrograde}
\end{equation}
where ``pro'' and ``ret'' stand for prograde (co-rotating) orbits and retrograde (counter-rotating) orbits, respectively.


\subsection{Stable circular orbits : away from the equatorial plane}
\label{sec-2.2}

We will now discuss some known results on off-equatorial circular orbits for massive objects in the Kerr BH background, that will be 
relevant for our analysis below. For such non-equatorial orbits, Carter's constant, $ Q \ne 0 $. To obtain the stable orbits on different 
planes for some fixed $ r $, we will follow an algorithm that is clearly described in \cite{Hughes1}.

\begin{itemize}
	\item[1.] Start with the prograde equatorial orbit which we specify as the most stable orbit. Calculate the values of 
	$ E^{pro} $ and $ L^{pro} $ for a given $ r $ using Eq.(\ref{eq.E-L-prograde}). In this case, $ Q=0 $.
	
	\item[2.] The inclination of the orbit is now changed by gradually decreasing the value of $ L $, keeping the value 
	of $ r $ fixed. Out of the three constants of motion ($ E,L,Q $), only one is varied independently. 
	The other two can be expressed as a function of the independent one. It is convenient to vary $ L $ independently 
	and express $ E $ and $ Q $ as functions of $ L $, for a fixed r. To obtain the analytical forms of $ E(r,L) $ and $ Q(r,L) $, 
	we need to solve the same set of equations, $ R(r) = R'(r) = 0 $. The solution yields \cite{Hughes1},\cite{Hughes2}
	\begin{equation}
		E(r,L) = \frac{a^2 L^2 (r-M) + r \Delta^2}{a L M (r^2 - a^2) \pm \Delta \sqrt{r^5 (r-3M) + a^4 r (r+M) + 
		a^2 r^2 (L^2 - 2M r + 2r^2)}} ~ \label{eq.E-non-equator}
	\end{equation}
	\begin{equation}
	Q(r,L) = \frac{\left[ (a^2 + r^2) E(r,L) - a L \right]^2}{\Delta} - \left[ r^2 + a^2 E^2(r,L) - 2 a L E(r,L) + L^2 \right] 
	\label{eq.Q-non-equator}
	\end{equation}
The above $ E(r,L) $ has two solutions corresponding to the `$ + $' and `$ - $' signs in the denominator of 
Eq.(\ref{eq.E-non-equator}). Depending on the given value of $ r $, only one of them is physically relevant. 
The expression within the square root in the denominator of Eq.(\ref{eq.E-non-equator}) goes to zero at some 
value of $ r$ ($= {\mathcal R}(a)$, say). For $ r<{\mathcal R}(a) $, the `$ - $' sign is valid and for $ r>{\mathcal R}(a) $, 
the `$ + $' sign is physical; (in general, $ {\mathcal R} (a) $ is found to be close to $ 2M $) \cite{Hughes2}.
	
	\item[3.]
	A circular orbit is therefore determined completely by specifying the values of $ r $ and $ L $. The radius of the 
	orbit is fixed by $ r $ and the orbit is ascertained by $ L $, which specifies by 
	how much it is inclined with respect to the equatorial plane. As described in \cite{Hughes1},\cite{Hughes2},\cite{Ryan}, 
	this inclination angle can be defined by
	\begin{equation}
		\cos i = \frac{L}{\sqrt{L^2 + Q}} \label{eq.inc-angle-cos}
	\end{equation}
	where $ i \in [0,\pi] $. Here, $ i < \frac{\pi}{2} $ representing co-rotational motion and $ i > \frac{\pi}{2} $ representing counter-rotation. 
	Another useful description of the inclination angle is \cite{Hughes3},
	\begin{equation}
	\theta_{inc} = \frac{\pi}{2} - D \theta_{min} \label{eq.inc-angle-theta}
	\end{equation}
where $ \theta_{min} $ represents the minimum angle of $ \theta $ obtained during the orbital motion \cite{Wilkins}
and $ D = \pm 1 $, ($ +1 $ for prograde orbits and $ -1 $ for retrograde motion). 
The value of $ \theta_{inc} $ ranges from $ 0 $ to $ \frac{\pi}{2} $ for prograde motion and from $ \frac{\pi}{2} $ to $ \pi $ 
for retrograde orbits. We can easily find the conversion relations between $ i $ and $ \theta_{inc} $, and it shows that the 
difference between $ i $ and $ \theta_{inc} $ is small, matching exactly for $ a = 0 $. 
	
	\item[4.]
	For the fixed $ r $, the decrement of the value of $ L $ is carried on until either it equals the value of $ L^{ret} $ 
(as per the second expression of Eq.(\ref{eq.E-L-retrograde})) maintaining the required stability condition $ R'' < 0 $, or it 
reaches a specific value $ L^{mb} $ corresponding to $ R''(r)=0 $. The retrograde equatorial orbit with $ L^{ret} $ is identified 
as the least stable orbit. On the other hand, orbits with $ L^{mb} $ corresponding to $ R''(r) = 0 $ are called marginally 
bound stable orbits. Further decrease in the value of $ L $ than $ L^{mb} $ results in $ R''(r) > 0 $ which represents 
unstable circular orbits. Since we are interested only in stable circular orbits, we will not consider such unstable ones.
		
	Figure \ref{fig.orbits} shows some representative non-equatorial, circular orbits for different values of the parameters. 
All the plots are drawn for $ r=9 $ (in units of $G=c=1$)  for illustration, as for this radius, both the prograde ($ L>0 $) as well as 
retrograde ($ L<0 $) stable orbits are possible. The inclination angles of the orbits ($ \theta_{inc} $), 
with respect to equatorial plane, are chosen to be different for the three plots drawn in figures \ref{fig.orbits30}, \ref{fig.orbits60}
and \ref{fig.orbits135}, with $ \theta_{inc}=\frac{\pi}{6},\frac{\pi}{3}~{\rm and}~\frac{3\pi}{4} $ respectively. 
The direction of rotation of the Kerr background is shown by a blue arrow-head on top of each sphere 
and the rotation of the stars on the surface of each sphere is directed along red arrow-heads in the circular orbits. 
From the direction of the arrow-heads, it is clear that the orbits with $ \theta_{inc}=\frac{\pi}{6} $ and $ \frac{\pi}{3} $ 
represent prograde orbits, whereas $ \theta_{inc}=\frac{3\pi}{4} $ indicates a retrograde orbit.
	
	\begin{figure}[H]
		\centering
		\begin{subfigure}{.3\textwidth}
			\includegraphics[width=\textwidth]{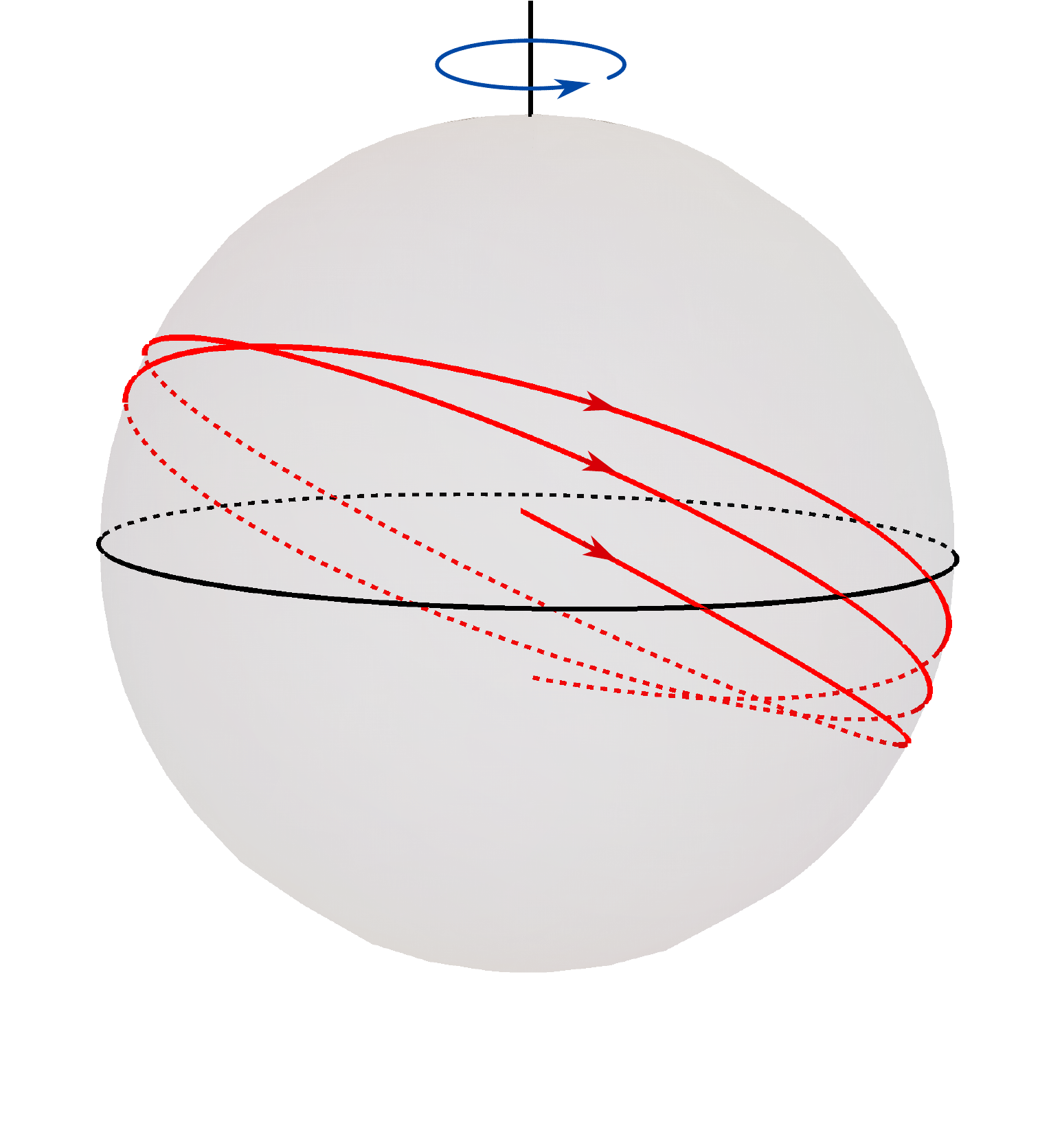}
			\caption{ $ \theta_{inc}= \pi/6 $, or, $ 30^\circ $ }
			\label{fig.orbits30}
		\end{subfigure}
		\begin{subfigure}{.3\textwidth}
			\includegraphics[width=\textwidth]{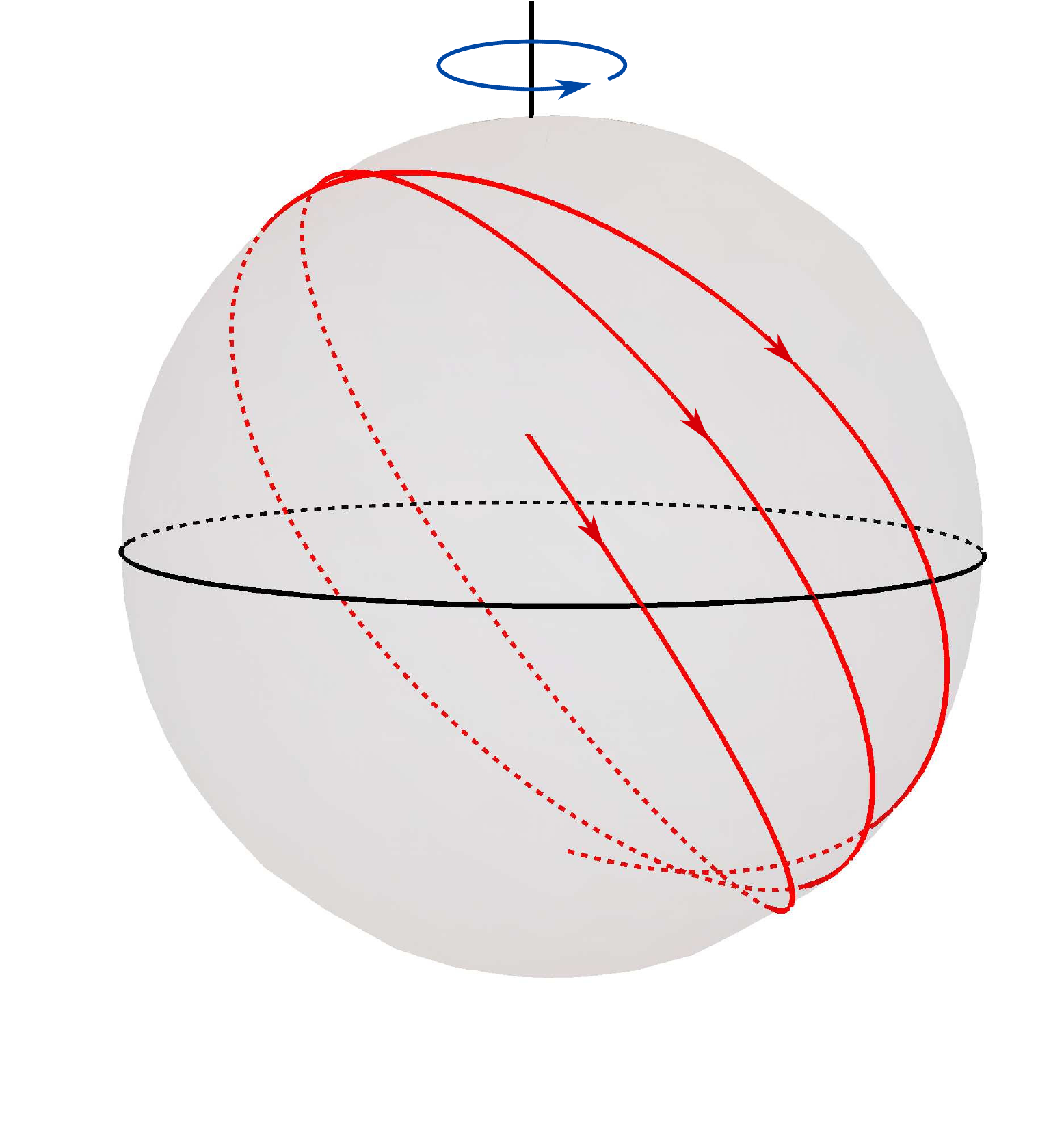}
			\caption{ $ \theta_{inc}= \pi/3 $, or, $ 60^\circ $ }
			\label{fig.orbits60}
		\end{subfigure}
		\begin{subfigure}{.3\textwidth}
			\includegraphics[width=\textwidth]{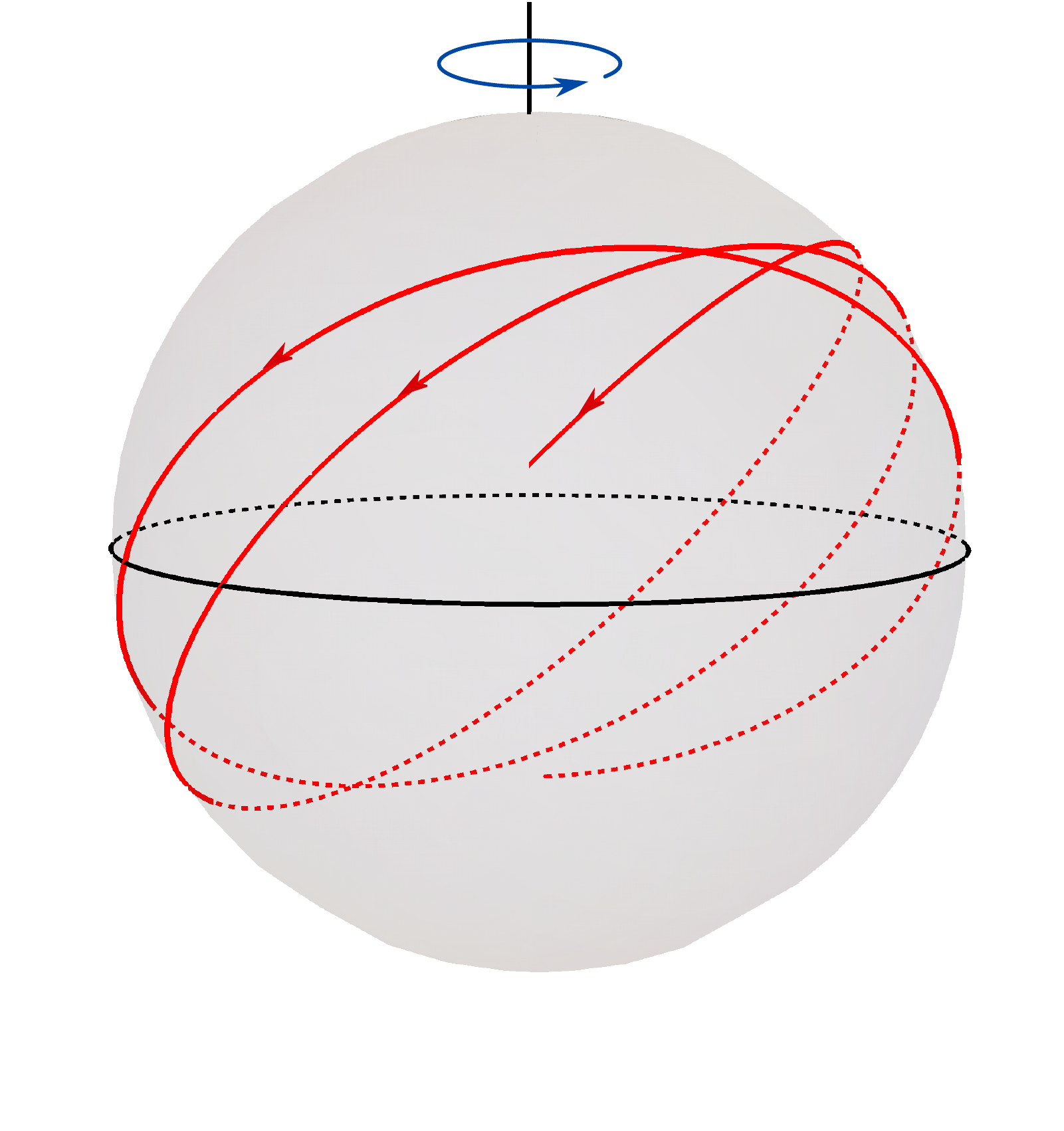}
			\caption{ $ \theta_{inc}= 3\pi/4 $, or, $ 135^\circ $ }
			\label{fig.orbits135}
		\end{subfigure}
		\caption{Representative orbits plot for $ r=9 $. (a) represents orbit with $ \theta_{inc}=\pi/6 $, (b) stands for 
		$ \theta_{inc}=\pi/3 $, and (c) indicates $ \theta_{inc}=3\pi/4 $. (a),(b) are prograde orbits, and (c) is retrograde. 
		The black circle in the middle of each sphere represents the equator. }
		\label{fig.orbits}
		
	\end{figure}
	
\end{itemize}
For convenience, let us now summarize the recipe of obtaining non-equatorial, stable, circular orbits in the Kerr background. 
First, we fix the radius ($ r $) of the orbit, and find out $ E^{pro} $ and $ L^{pro} $ for the equatorial plane. 
Then we decrease the value of $ L $ which changes the inclination of the orbit, and find out $ E(r,L) $ and $ Q(r,L) $ 
corresponding to that plane, subject to the condition, $ R''(r) < 0 $. Therefore, $ L $ is lowered until either it attains the 
value of $ L^{ret} $ corresponding to the retrograde orbit on the equatorial plane, or we obtain $ L^{mb} $ corresponding to 
$ R''(r) = 0 $. The orbits with $ R''(r) = 0 $ represent marginally bound stable orbits, so that values of $ L $ less than $ L^{mb} $ 
satisfying $ R''(r) > 0 $ produce unstable orbits.


\section{Tidal Effects in Non-equatorial Orbits}
\label{sec-3}

Let us consider a star rotating around a Kerr BH in a stable, circular trajectory. We want to find out the tidal disruption limit 
of the star due to the influence of the BH. As  mentioned, we will follow the formalism developed in \cite{Ishii-Kerr}, and the 
formulae which do not need modifications in our calculation will be referred to from that paper. 


\subsection{Formulation}
\label{sec-3.1}

Let us write the hydrodynamic equation of the fluid star in FNC as 

\begin{equation}
	\rho \frac{\partial v_i}{\partial \tau} + \rho v^j \frac{\partial v_i}{\partial x^j} = - \frac{\partial P}{\partial x^i} - 
	\rho \frac{\partial (\phi + \phi_{\text{tidal}})}{\partial x^i} +\rho \left[ v^j \left( \frac{\partial A_j}{\partial x^i} - 
	\frac{\partial A_i}{\partial x^j} \right) - \frac{\partial A_i}{\partial \tau} \right] \label{eq.hydrodynamic}
\end{equation}
where $ \rho $ is the fluid density, $ v^i $ is the three-velocity of the fluid ($ \frac{dx^i}{d\tau} $), $ P $ is the fluid pressure, 
$ A_i $ is a vector potential associated with the gravito-magnetic force \cite{gravito-magnetic}, $ \phi $ is the Newtonian 
self-gravitational potential of the star, and $ \phi_{\text{tidal}} $ is the tidal potential produced by the Kerr BH. 
In terms of FNC \{$ x^0 (=\tau), x^1, x^2, x^3 $\}, $ \phi_{\text{tidal}} $ is given by \cite{Ishii-Kerr}

\begin{eqnarray}
\phi_{\text{tidal}} &=& - \frac{1}{2} (g_{00} + 1) \nonumber\\
&=& - \frac{1}{4} G_{00,ij} x^i x^j - \frac{1}{12} G_{00,ijk} x^i x^j x^k - \frac{1}{48} G_{00,ijkl} x^i x^j x^k x^l + O(x^5) \nonumber\\
&=& \frac{1}{2} C_{ij} x^i x^j + \frac{1}{6} C_{ijk} x^i x^j x^k + \frac{1}{24} \left[ C_{ijkl} + 
4 C_{\left(ij\right.} C_{\left.kl\right)} - 4 B_{\left(kl|n|\right.} B_{\left.ij\right)n} \right] x^i x^j x^k x^l + O(x^5) 
\label{eq.phi-tidal}
\end{eqnarray}
where we have defined
\begin{equation}
C_{ij} = R_{0i0j}, ~~~ C_{ijk} = R_{0\left(i|0|j;k\right)}, ~~~ C_{ijkl} = R_{0\left(i|0|j;kl\right)}, ~~~ B_{ijk} = R_{k\left(ij\right)0} 
\label{eq.Cij}
\end{equation}
The vector potential $ A_i $ is defined as
\begin{equation}
A_i = \frac{2}{3} B_{ijk} x^i x^j \label{eq.vector-potential}
\end{equation}
Here, the symbols `;' and `,' in between the indices of $ R,g, $ etc. have the usual meaning as the covariant derivative and 
the ordinary (partial) derivative respectively. Moreover, $  R_{0\left(i|m|j;kl\right)} $ indicates a summation over all the permutations 
of the indices $ i,j,k $ and $ l $, keeping $ m $ fixed at its position and then division by the total number of such permutations. 
The gravito-magnetic term in the hydrodynamic equation (Eq.(\ref{eq.hydrodynamic})) is important as the magnitude of this term 
becomes as large as the fourth order terms in $ \phi_{\text{tidal}} $ for the co-rotational velocity field of the star. The potential 
due to the self-gravity of the star ($ \phi $) satisfies the Poisson's equation of Newtonian gravity, given by
\begin{equation}
	\nabla^2 \phi = 4 \pi \rho \label{eq.poisson}
\end{equation}
where $ \rho $ is the mass density profile of the star. We are considering a co-rotational star which is static in the tilde frame defined as
\begin{equation}
\tilde{x}^1 = x^1 \cos \Psi + x^3 \sin \Psi~,~~
\tilde{x}^2 = x^2~,~~
\tilde{x}^3 = - x^1 \sin \Psi + x^3 \cos \Psi 
\label{eq.tilde-coord}
\end{equation}
where, the angle $ \Psi $ is associated with the parallel transportation of the Fermi normal frame (Eqs.(113)-(116) and Eqs.(121)-(124) of \cite{Ishii-Kerr}). In the tilde frame, fluid velocity is zero. But in the Fermi Normal frame, it is given by\footnote{We note that there is a possible typographic error in the expression of $ v^i $ in Eq.(167) of \cite{Ishii-Kerr}. 
	An extra $ \Omega $ should be multiplied in front of the square bracket in the right hand side of that equation. 
	We have written the expression inclusive of this factor.}
\begin{equation}
v^i = \Omega [-\{x^3 - x_c \sin\Psi\},0,\{x^1 - x_c \cos\Psi\}] \label{eq.vel-field}
\end{equation}
where $ \Omega = d\Psi / d\tau $ and $ x_c $ is a correction term which arises due to the fact that in the tilde frame, in presence of the third order term of the tidal potential or the gravito-magnetic effects, the center of mass of the star is deviated from the origin. By examining the magnitudes of the components of the tidal tensors 
(Eq.(\ref{eq.Cij})) it can be understood that, in the tilde frame, the $ \tilde{x}^3 $ component of the position vector of the center of mass of the star is small enough to be neglected compared to the $ \tilde{x}^1 $ and $ \tilde{x}^2 $ components. It suggests that the center of mass is shifted from origin mostly in the $ \tilde{x}^1-\tilde{x}^2 $ plane. Moreover, since the tilde frame is rotating about the $ x^2 $ axis of the Fermi Normal frame, the $ \tilde{x}^2 $ component does not appear in the velocity expression (Eq.(\ref{eq.vel-field})). Therefore, only the $ \tilde{x}^1 $ component has been considered as $ x_c $.

Now, substituting for $ v^i $ in Eq.(\ref{eq.hydrodynamic}), integrating it, and then transforming to $ \tilde{x}^i $, 
the hydrodynamic equation becomes
\begin{equation}
\frac{\Omega^2}{2} \left[ (\tilde{x}^1 - x_g)^2 + (\tilde{x}^3)^2 \right] + \frac{d\Omega}{d\tau} \tilde{x}^3 x_c = 
h + \phi + \phi_{\text{tidal}} + \phi_{\text{mag}} + C  \label{eq.basi-eq-2}
\end{equation}
where $ x_g = 2 x_c $, $ C $ is an integration constant, $ \phi_{\text{mag}} $ is the scalar potential due to gravito-magnetic 
effects arising from the term involving $ A_i $ (computed from the last term on the right hand side of 
eq.(\ref{eq.hydrodynamic})), and $ h = \int \frac{dP}{\rho} $. The correction term $ x_c $ may depend on $ \theta $ for nonequatorial circular orbits but the dependence being too small, the term including $ \frac{dx_c}{d\theta} $ has been neglected. It is important to note that there is an extra 
term in Eq.(\ref{eq.basi-eq-2}) involving $ d\Omega/d\tau $ which is absent in the corresponding equation of \cite{Ishii-Kerr} (Eq.(168)
of that paper).
This is because $ \Omega $ ($ = d\Psi/d\tau $) depends only on $ r $ for equatorial orbits, and for non-equatorial orbits, 
it depends on both $ r $ and $ \theta $. So in case of circular motion in an equatorial orbit, 
it is a constant. For circular non-equatorial orbits, $ \Omega  = \Omega(\theta) $, or $ d\Omega/d\tau = (d\Omega/d\theta)\dot{\theta} $. 
The exact expression for $ d\Psi/d\tau$ is given by (Eq.(126) of \cite{Ishii-Kerr})
\begin{equation}
\Omega = \frac{d\Psi}{d\tau} = \frac{\sqrt{K}}{\Sigma} \left( \frac{E(r^2+a^2)-a L}{r^2+K} + \frac{a(L-aE\sin^2\theta)}{K-a^2\cos^2\theta} \right) \label{eq.Omega}
\end{equation}
where $ K = (L-aE)^2+Q $ is a constant, known as the modified Carter's constant.
Equations (\ref{eq.poisson}) and (\ref{eq.basi-eq-2}) constitute the basic equations for our analysis.


\subsection{Methodology}
\label{sec-3.2}

We start with the polytropic equation of state for the star given as
\begin{equation}
P = \kappa \rho^\Gamma, ~~~ \text{where} ~~ \Gamma = 1 + \frac{1}{n}, ~~~ \text{so that}, ~~~ 
h = \kappa (n + 1) \rho^{\frac{1}{n}} \label{20}
\end{equation}
where $ \kappa $ is called polytropic constant and $ n $ is the polytropic index. Eqs.(\ref{eq.poisson}) 
and (\ref{eq.basi-eq-2}) are solved together numerically as coupled equations. To obtain numerical convergence, 
we convert the equations into dimensionless ones. Therefore, writing the coordinates as $ \tilde{x}^i = p q^i $, where 
$ p $ is a constant with dimension of length and $ q^i $'s are dimensionless coordinates, Eqs.(\ref{eq.poisson}) 
and (\ref{eq.basi-eq-2}) respectively become

\begin{equation}
\nabla_q^2 \bar{\phi} = 4 \pi \rho \label{eq.poissn-new}
\end{equation}
\begin{equation}
\frac{\Omega^2}{2} p^2 \left[ (q^1 - q_g)^2 + (q^3)^2 \right] + p^2 \frac{d\Omega}{d\tau} q^3 q_c = h(\rho) + p^2 \left(\bar{\phi} + \bar{\phi}_{\text{tidal}} + \bar{\phi}_{\text{mag}} \right) + C  \label{eq.basic-eq-2-new}
\end{equation}
where $ \nabla_q $ is the Laplacian operator in terms of $ q^i $ coordinates, $ q_g = p^{-1} x_g $, $ \bar{\phi} = p^{-2}\phi $, 
$ \bar{\phi}_{\text{tidal}} = p^{-2}\phi_{\text{tidal}} $, and $ \bar{\phi}_{\text{mag}} = p^{-2}\phi_{\text{mag}} $. 
The numerical recipe to obtain the tidal disruption limit or Roche limit of the star is \cite{Ishii-Kerr}
 \begin{itemize}
 	\item[1.] Consider a spherically-symmetric density profile, $ \rho(q^i) $, of the polytrope as a trial function, 
 	with a specific value of $ n $ or $ \Gamma $. This is a solution of the corresponding Lane-Emden equation for the given
 	value of $ n $. In the coordinate system ($ \tilde{x}^1,\tilde{x}^2,\tilde{x}^3 $) or ($ q^1,q^2,q^3 $), the $ r $-direction (i.e. the BH direction) varies with the angular position($ \theta $) of the star as
 	\begin{equation}
 	q^2 = \frac{a \cos (\theta )}{r \sqrt{\frac{K-a^2 \cos ^2(\theta )}{K+r^2}}} q^1
 	\label{BH-direction}
 	\end{equation}
 	and it is exactly aligned with the $ q^1 $-direction on the equatorial plane.    
 	
 	\item[2.] Put $ \rho $ in the right hand side of Poisson's equation (\ref{eq.poissn-new}), and solve it numerically to 
 	obtain $ \bar{\phi} $. We use the cyclic reduction method to solve the corresponding matrix equations using 
 	Dirichlet boundary conditions. Moreover, we consider a cubic volume with $ 101 \times 101 \times 101 $ 
 	grid points to obtain the solution. The cubic boundary is set at 50 grid points away from the center grid point. The size of the star is assumed to be smaller than the size of the cube so that on the surface of the star, the potential can be approximated to $ -\int_{0}^{R_0} \frac{\rho ~ d^3q}{r_q} $, 
 	where $ r_q $ is the radial distance from the center, and $ R_0 $ is the average radius of the star. Since $ \rho $ is 
 	anyway zero outside the star, we can re-write the integral as $ -\int_{cube} \frac{\rho ~ d^3q}{r_q} $ to simplify 
 	the computation \cite{Karmakar-Tapo}.
 	
 	\item[3.] The next step is to find out different constants of the problem using the recently obtained $ \bar{\phi} $. 
 	There are seven free constants to be determined, namely $ M,~ a,~ \kappa,~ p,~ q_g,~ C $ and 
 	$ \rho_c $(central density of the star). All the calculations are performed in units, $ c=G=M=1 $. 
 	This fixes the value of $ a $ to be chosen in the range $ -1 \le a \le 1 $. 
 	The value of $ \kappa $ is determined from the average radius ($ R_0 $) 
 	of the star following the relation
 	
 	\begin{equation}
 	R_0 = \left[ \frac{(1+n)\kappa\rho_c^{(1-n)/n}}{4 \pi} \right]^{\frac{1}{2}}\xi_1 \label{eq.R-not}
 	\end{equation}
 	where $ \xi_1 $ is the Lane-Emden parameter at the surface (with known values $\pi$, $6.89685 $ for 
 	$ n= 1$ and $3$, respectively). So the remaining three constants are $ p,~ q_g $ and $C$. The 
 	central density ($ \rho_c $) of the star is chosen to be a maximum, i.e. $ \left.\rho\right\vert_{(0,0,0)} = \rho_c $ 
 	and $ \left.\frac{\partial\rho}{\partial q^1}\right\vert_{(0,0,0)} = 0 $. Moreover, the stellar surface is fixed at $ (q_s^1,q_s^2,0) $ along the $ r $-direction such that $ \left.\rho\right\vert_{(q_s^1,q_s^2,0)} = 0 $, where $ q_s^1 $ and $ q_s^2 $ satisfy Eq.(\ref{BH-direction}).
 	These three conditions together determine $ q_g $, $ C $ and $ p $. On the equatorial plane, $ (q_s^1,q_s^2,0) $ is considered at 40 grid points away from the center on the $ q^1 $ axis. For a non-equatorial position, $ (q_s^1,q_s^2,0) $ is chosen 
 	at the same distance but along the $ r $-direction and an appropriate (nearest) grid point is used for numerical convergence.
 	
 	\item[4.] Once all the constants are determined, a new density profile is evaluated from $ h(\rho) $ using Eq.(\ref{eq.basic-eq-2-new}).
 	
 	\item[5.] Now substitute this new density profile into the r.h.s. of Eq.(\ref{eq.poissn-new}) and obtain an updated $ \bar{\phi} $.
 	
 \end{itemize}
We continue steps $ 2 $ -- $ 5 $ repeatedly until sufficient convergence is obtained. Now, the Roche limit is determined from 
a critical value of the central density.  We start the numerical computation with a sufficiently large $ \rho_c $, and gradually decrease it 
to smaller values until we obtain the critical value ($ \rho_{\text{crit}} $) for which the star just remains in a stable configuration. 
This is the condition when the binding self-gravity of the star is just enough to balance the disruptive tidal effects at the surface. 
At this limit, the star surface at ($ q_s^1,q_s^2,0 $) begins to form a cusp. Therefore, at Roche limit \cite{Karmakar-Tapo}
\begin{equation}
	\left.(\hat{r}\cdot\nabla_q\rho)\right\vert_{(q_s^1,q_s^2,0)} = 0 \label{eq.Roche-condition}
\end{equation}
where $\hat{r}$ is the unit vector in the $r$-direction.
Beyond the Roche limit, i.e., stars with $ \rho_c < \rho_{\text{crit}} $ are tidally disrupted and the corresponding 
density contours at the surface start to break. Then $ \rho_{\text{crit}} $ is used to define a dimensionless quantity, 
$ \xi_{\text{crit}} = \frac{\Omega^2}{\pi \rho_{\text{crit}}} $ (following \cite{Fishbone}, this is the ratio of the tidal force to the 
force due to self gravity at the tidal disruption limit). Therefore, stars with 
$ \xi < \xi_{\text{crit}} $ will be stable against tidal disruption.


\subsection{Results and Analysis}
\label{sec-3.3}

In this subsection, we describe the numerical results of tidal effects away from the equatorial plane in the Kerr BH background. 
Here, all the calculations are performed in units $ c=G=M=1 $. We have chosen $ a=0.9 $ throughout our analysis. 
We have calculated $ \xi_{\text{crit}} $ for two values of $ n $, viz. $ n=1 $ and $ 3 $, which correspond to
 $ \Gamma = 2 $ and $ \frac{4}{3} $ respectively. It is worth pointing out that $n=1$ corresponds to a 
highly magnetized white dwarf, whereas $n=3$ corresponds to the white dwarf equation of state with relativistic degenerate
electrons, without a magnetic field. 
 
\begin{figure}[H]
	\centering
	\begin{subfigure}{.3\textwidth}
		\includegraphics[width=\textwidth]{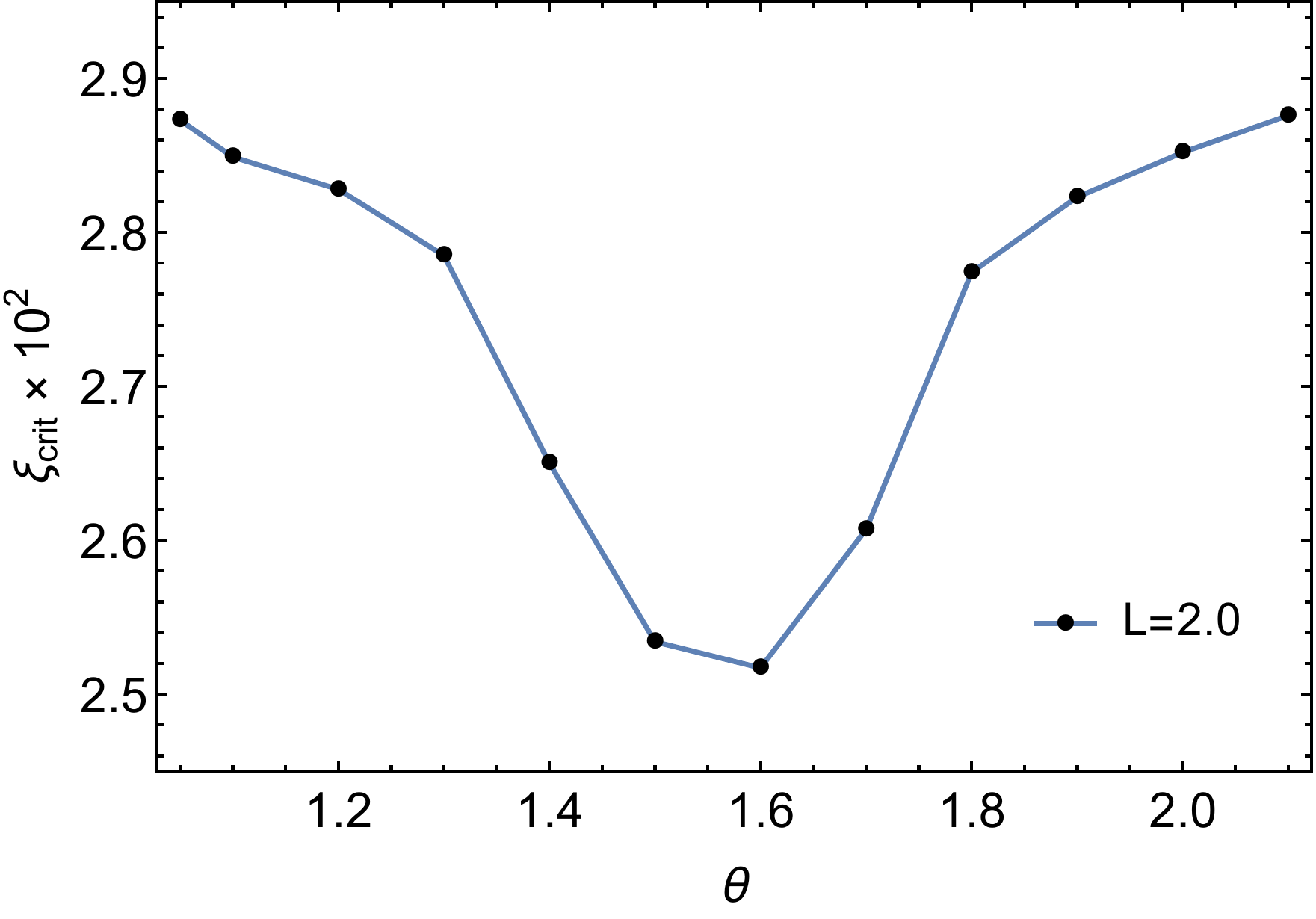}
		\caption{ $ r=3 $, $ n=1 $ }
	\end{subfigure}
	\begin{subfigure}{.3\textwidth}
		\includegraphics[width=\textwidth]{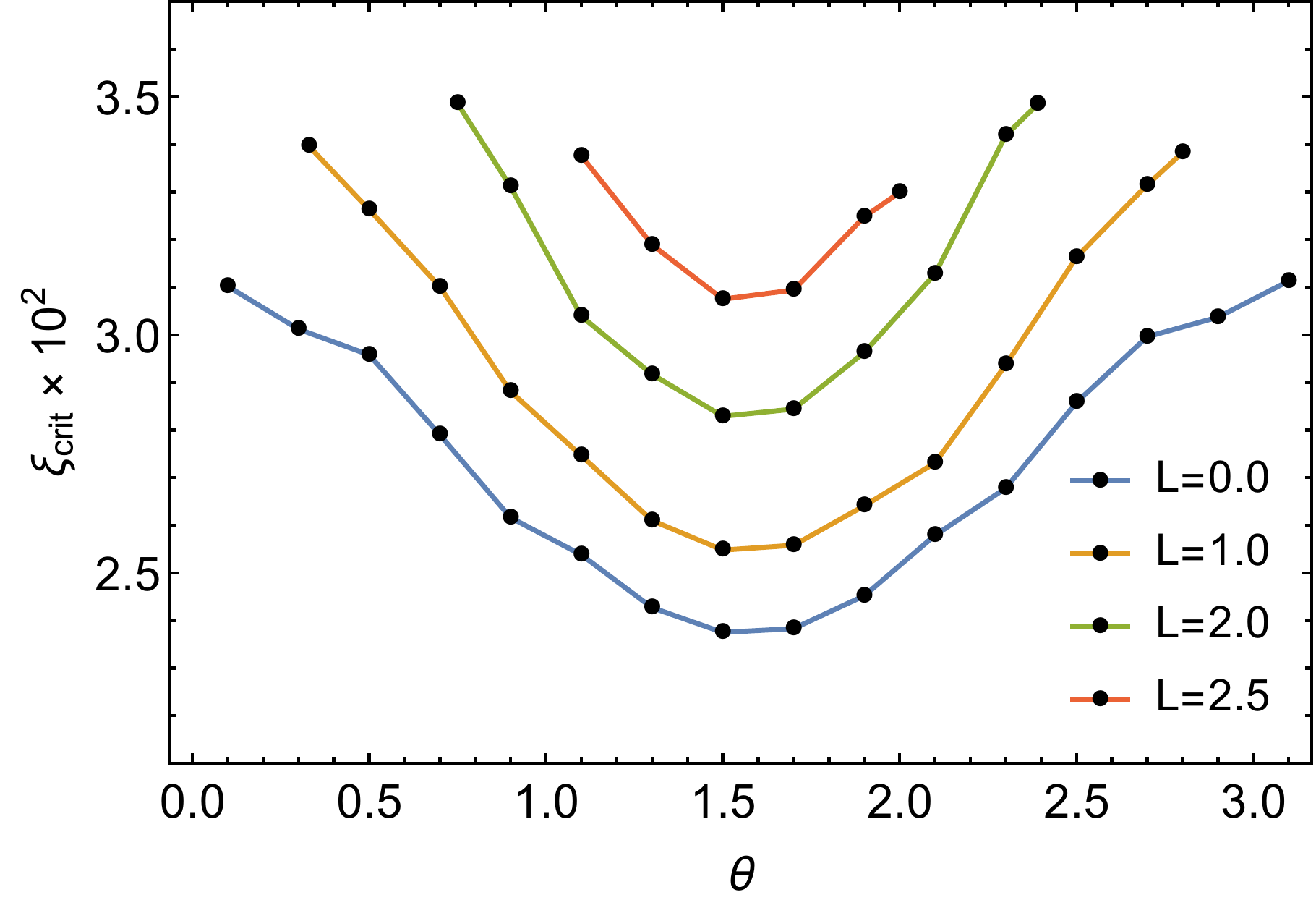}
		\caption{ $ r=6 $, $ n=1 $ }
	\end{subfigure}
    \begin{subfigure}{.3\textwidth}
	\includegraphics[width=\textwidth]{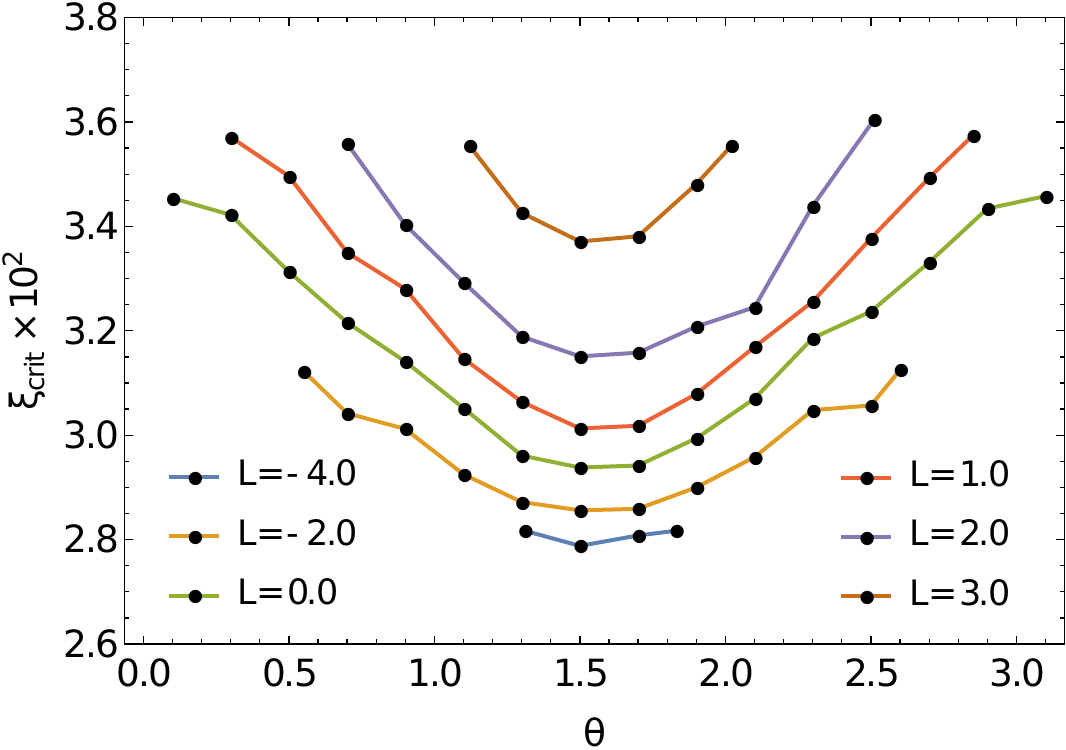}
	\caption{ $ r=9 $, $ n=1 $ }
    \end{subfigure}
	\caption{Plots of $ \xi_{\text{crit}} $ (enhanced by a factor of $10^2$) vs $ \theta $ for $ n=1 $. A single curve is represented by a specific value of $ L $. (a) represents the plot for $ r=3 $, $ L=2.0 $; (b) shows plots for $ r=6 $, $ L=0.0,1.0,2.0,3.0 $; and (c) indicates plots for $ r=9 $, $ L=-4.0,-2.0,0.0,1.0,2.0,3.0 $. }
	\label{fig.Xi-vs-th}
	
\end{figure}

Figure (\ref{fig.Xi-vs-th}) shows the variation of $ \xi_{\text{crit}} $ as a function of $ \theta $ for different values of 
$ r $ and $ L $. The allowed range of $ L $ depends on $ r $ for circular orbits to be stable. In case of $ r=3,6 $ and $ 9 $, 
it is found that, $ 1.8278 \le L \le 2.1883 $, $ -0.5889 \le L \le 2.7943 $ and $ -4.1699 \le L \le 3.3047 $ respectively. 
The negative values of $ L $ represent retrograde orbits, and the positive values stand for prograde orbits. Again, 
the values of $ \theta $ are also bounded for a specific $ L $. When the value of $ L $ equals either $ L^{pro} $ 
or $ L^{ret} $ (if possible), i.e. on the equatorial plane, we know $ \theta = \pi/2 $. As the value of $ L $ deviates from 
its equatorial value, the allowed range of $ \theta $ starts broadening with an increase of the difference 
between $ \theta_{max} $ and $ \theta_{min} $. For orbits having inclination angles ($ \theta_{inc} $)
nearly equal to $ \pi/2 $ with 
respect to equator, $ \theta $ ranges from $ 0 $ to $ \pi/2 $. 

From Fig.(\ref{fig.Xi-vs-th}) we observe that $ \xi_{\text{crit}} $, for a fixed $ L $, has a higher value at 
$ \theta_{min} $ or $ \theta_{max} $, and is minimum at $ \theta = \pi/2 $. Moreover, as we increase the 
value of $ L $ from $ L^{ret} $ to $ L^{pro} $, the magnitude of $ \xi_{\text{crit}} $ increases. Therefore, 
co-rotating stars are more stable than the corresponding counter-rotating ones, and on a particular orbit, 
stability is maximum at the two extreme points of their orbits (at $ \theta_{min} $ or $ \theta_{max} $), 
while it is minimum at the equator.

\begin{figure}[H]
	\centering
	\begin{subfigure}{.3\textwidth}
		\includegraphics[width=\textwidth]{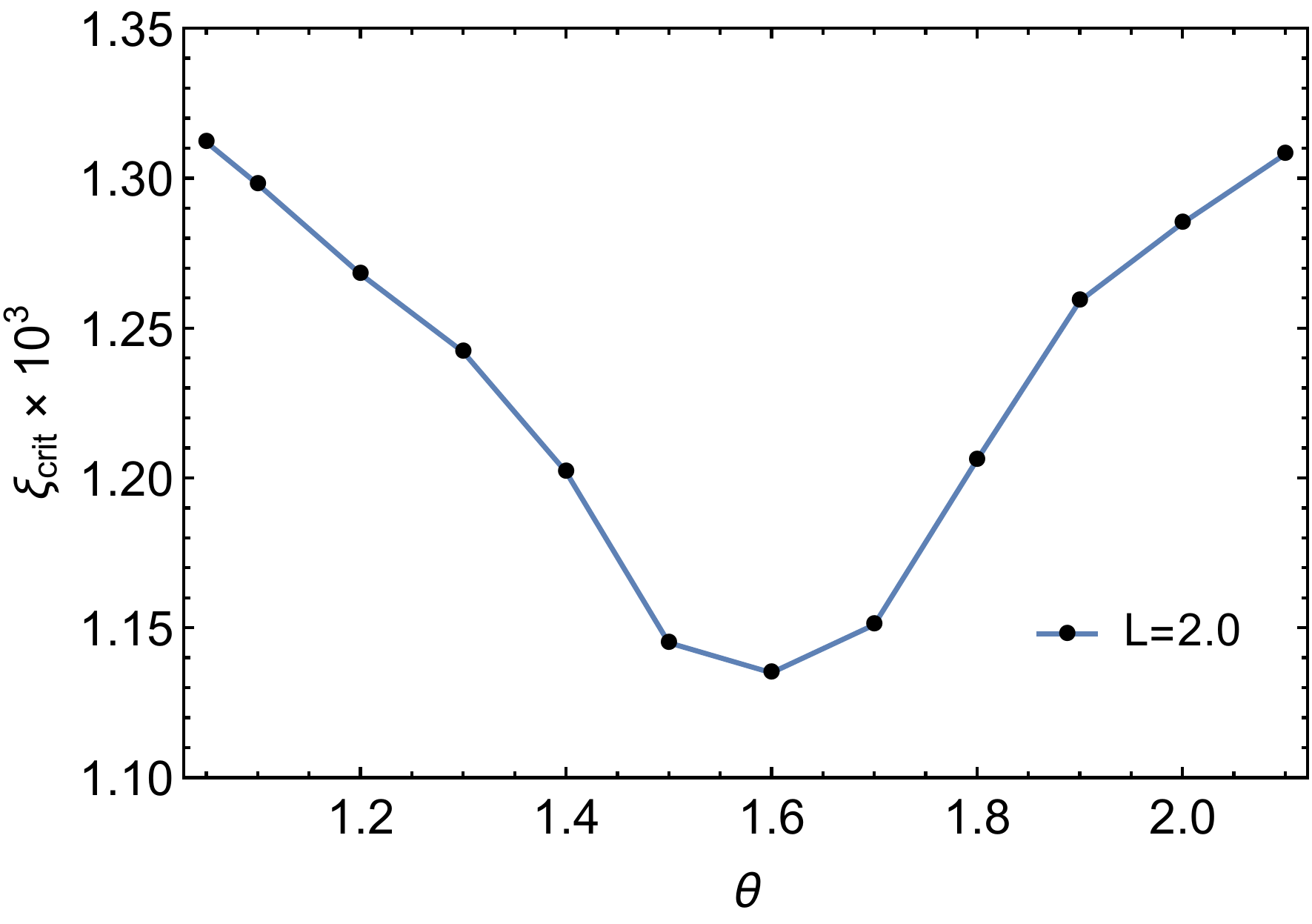}
		\caption{ $ r=3 $, $ n=3 $ }
	\end{subfigure}
	\begin{subfigure}{.3\textwidth}
		\includegraphics[width=\textwidth]{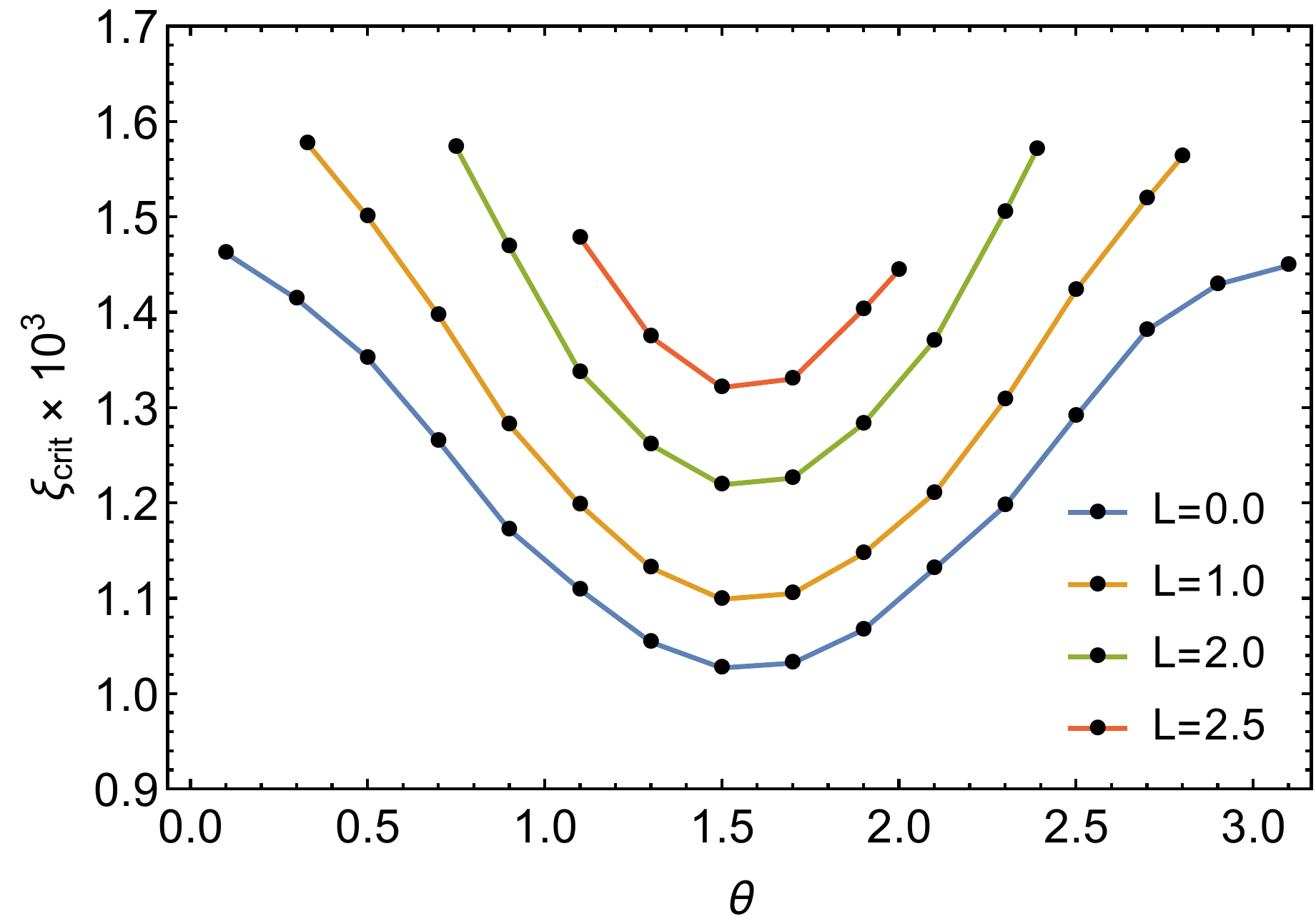}
		\caption{ $ r=6 $, $ n=3 $ }
	\end{subfigure}
	\begin{subfigure}{.3\textwidth}
		\includegraphics[width=\textwidth]{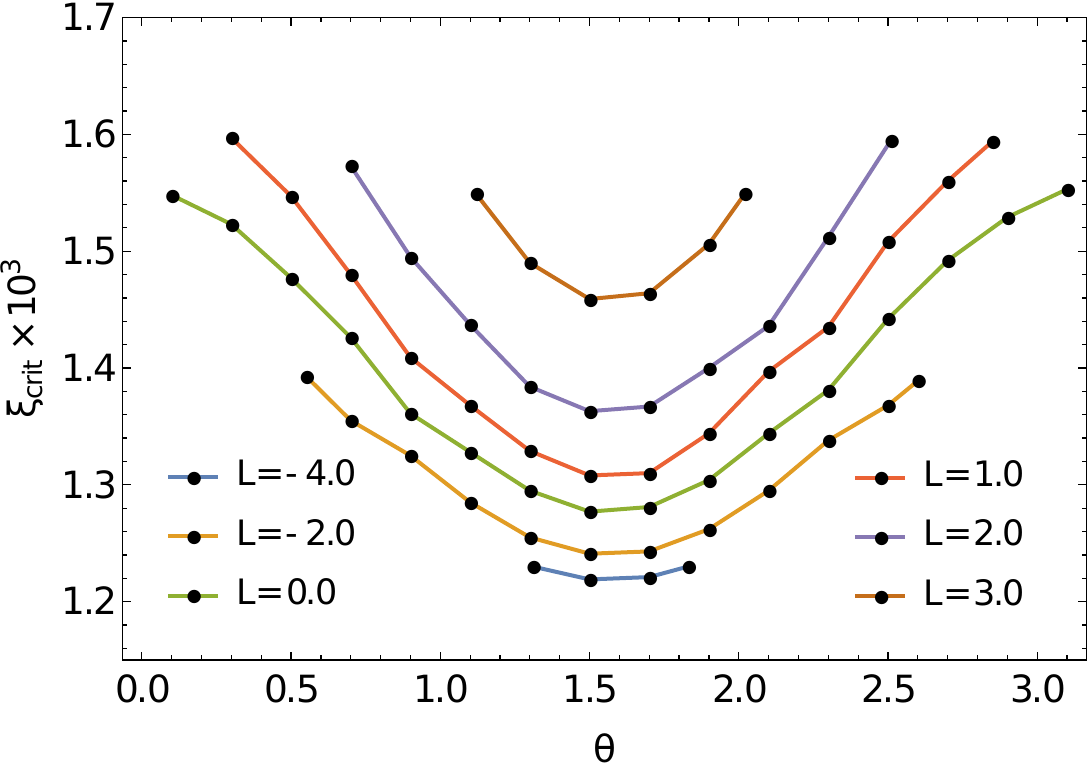}
		\caption{ $ r=9 $, $ n=3 $ }
	\end{subfigure}
	\caption{Plots of $ \xi_{\text{crit}} $ vs $ \theta $ (enhanced by a factor of $10^3$) for $ n=3 $. 
	The values of the parameters are the same as the $ n=1 $ case. }
	\label{fig.Xi-vs-th-n3}
	
\end{figure}

Figure (\ref{fig.Xi-vs-th-n3}) shows similar plots for $ n=3 $, taking the same values of $ L $ and $ \theta $ 
as above. The nature of the plots are almost similar to the $ n=1 $ case, which indicates that the $ \xi_{\text{crit}}-\theta $
 behavior is a generic feature of the orbits of the stars, and it does not depend on the equation of state parameter of stars. 
 An important point to note here is that the magnitudes of $ \xi_{\text{crit}} $ for $ n=1 $ case is approximately 
 $ 20 $ times larger than the corresponding magnitudes for $ n=3 $ case.

Therefore, we see that $ \xi_{\text{crit}} $ is sensitive to the equation of state of the star, and a star with 
higher $ n $ is less stable against tidal disruption than a star with lower $ n $. This is also in agreement with the equatorial
plane analysis of \cite{Ishii-Kerr}.

\begin{figure}[H]
	\centering
	\begin{subfigure}{.45\textwidth}
		\includegraphics[width=\textwidth]{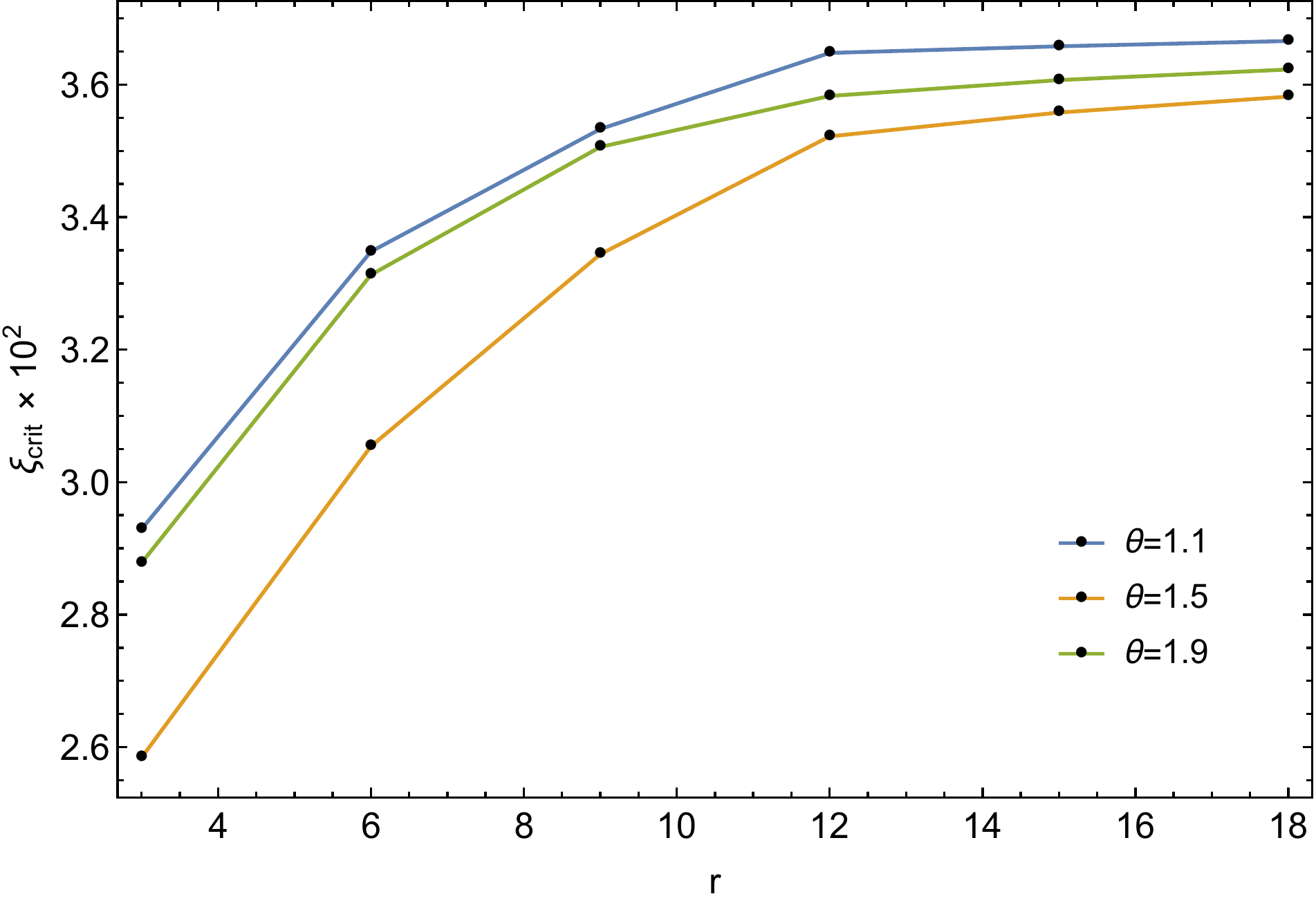}
		\caption{ $ n=1 $ }
	\end{subfigure}
	\begin{subfigure}{.45\textwidth}
	\includegraphics[width=\textwidth]{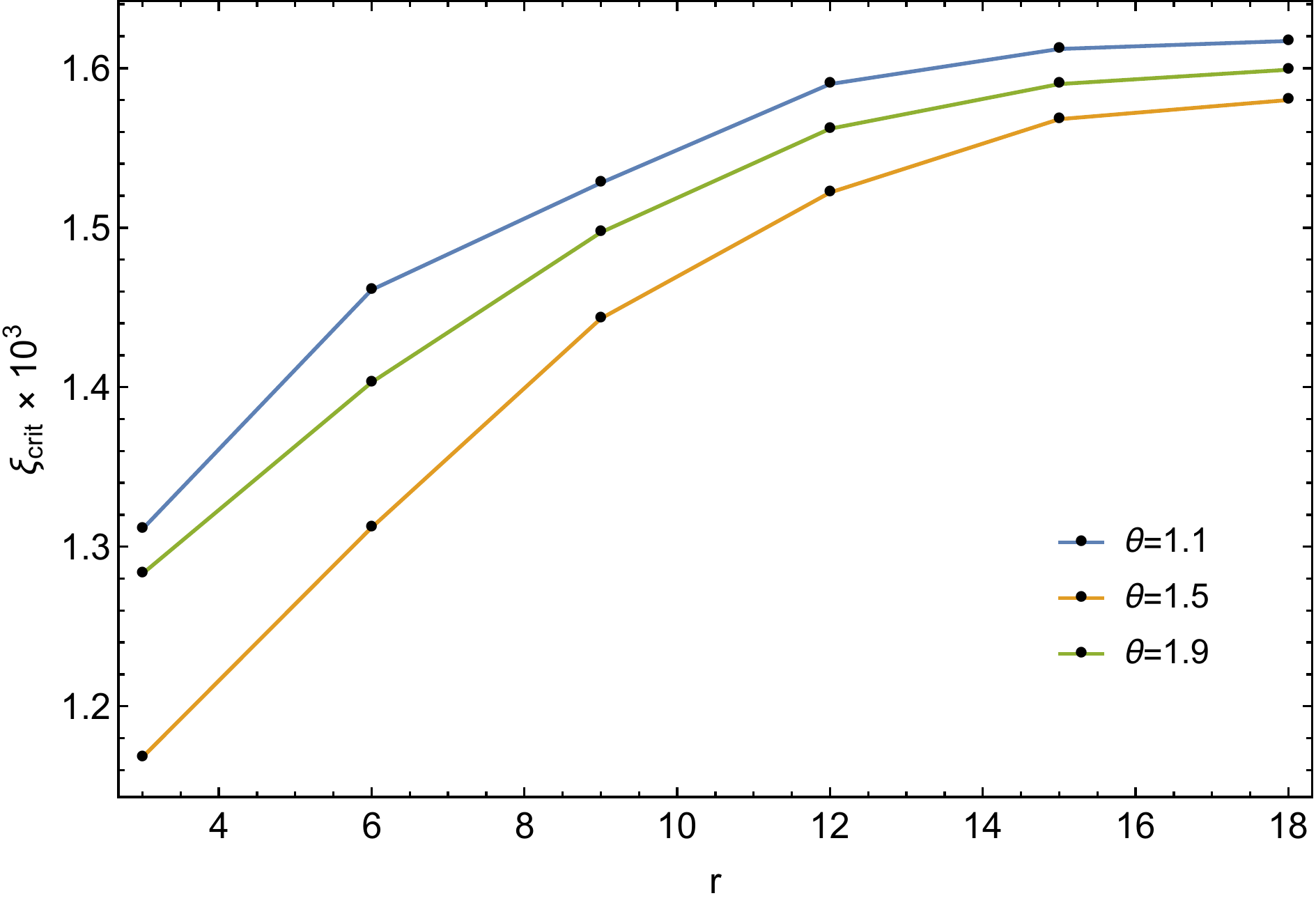}
	\caption{ $ n=3 $ }
	\end{subfigure}
	\caption{Variation of $ \xi_{\text{crit}} $ as a function of $ r $ (enhanced by a factor of $10^2$ for $n=1$ and $10^3$ for $n=3$). 
	We have chosen $ M=1 $. The three curves shows plots for three values of $ \theta $, viz. $ \theta=1.1,1.5,1.9 $. (a) represents the plot for $ n=1 $, and (b) stands for $ n=3 $. The nature of the curves are similar for both the values of $ n $ but their magnitudes are different, as stated earlier.}
	\label{fig.Xi-vs-r}
	
\end{figure}

Next, in Fig.(\ref{fig.Xi-vs-r}), we have shown how $ \xi_{\text{crit}} $ varies with  $ r $ for a given inclination 
angle $ \theta_{inc} $, i.e., on a fixed orbit. For simplicity, we have chosen a plane having $ \theta_{inc}=\pi/6 $, 
and on the same plane, we consider three different angular positions, $ \theta=1.1,1.5 $ and $ 1.9 $ to obtain the plots. 
From this figure it is again clear that $ \xi_{\text{crit}} $ has the lowest value on the equator i.e., $ \theta=\pi/2 $ and 
increases as the star moves away from the equator towards the extreme points of its orbit, i.e., the turning points 
on a specific orbit. The difference of Roche limit at various angles on a fixed orbit is more prominent when 
the star is closer to the BH and it reduces as $ r $ increases. Therefore, stars which are far away from the BH will 
remain fairly stable against tidal disruption for any orbit they rotate on. And as the star-BH distance 
reduces, stars (with the same mass and stellar radius) 
on or near the equatorial plane will start getting disrupted more easily than those off the equatorial plane.

\begin{figure}[h]
	\centering
	\begin{subfigure}{.32\textwidth}
		\includegraphics[width=\textwidth]{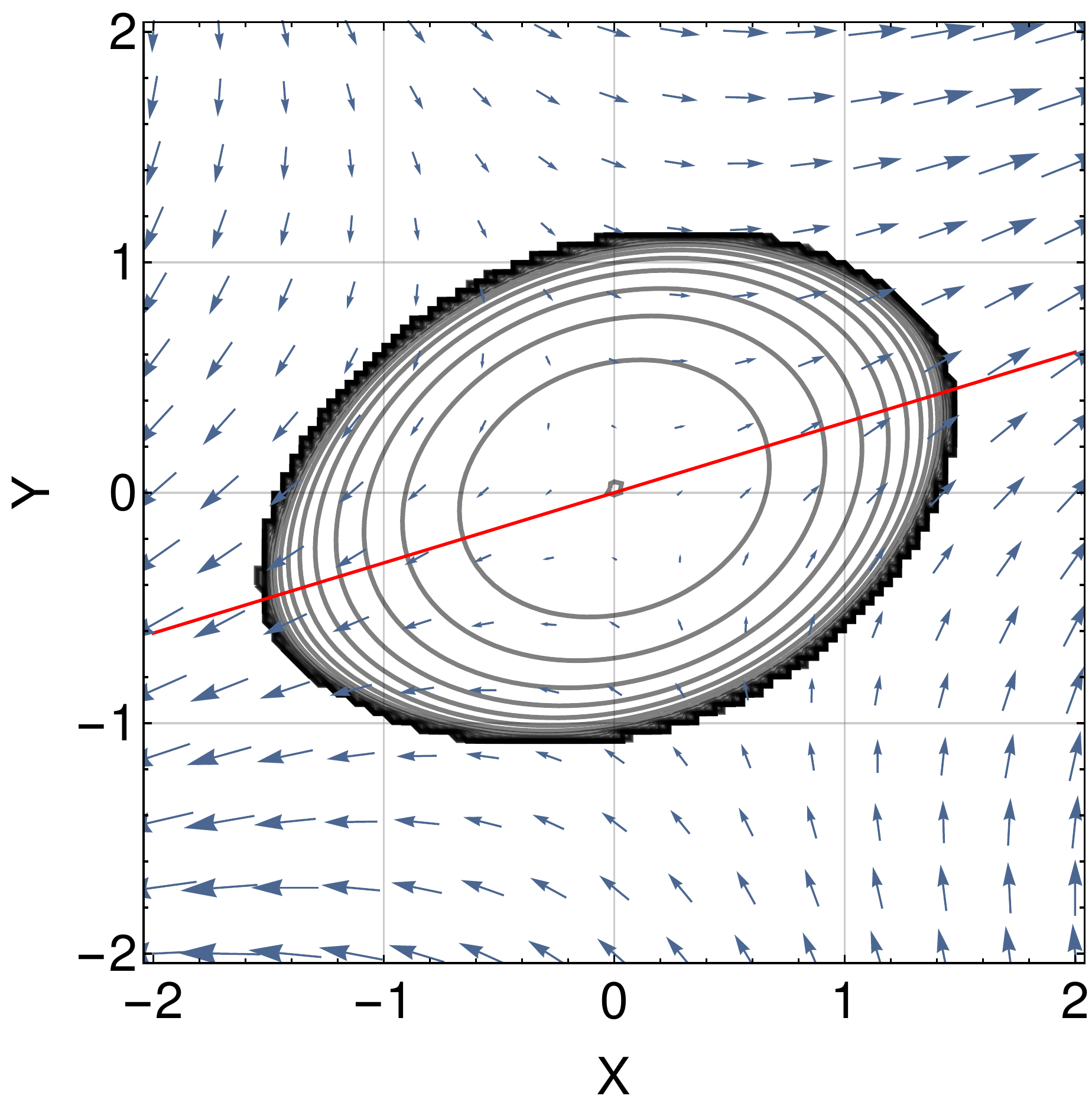}
		\caption{ $ \theta=0.5 $, $ n=1 $ }
		\label{fig.density-plots-n1-a}
	\end{subfigure}
	\begin{subfigure}{.32\textwidth}
		\includegraphics[width=\textwidth]{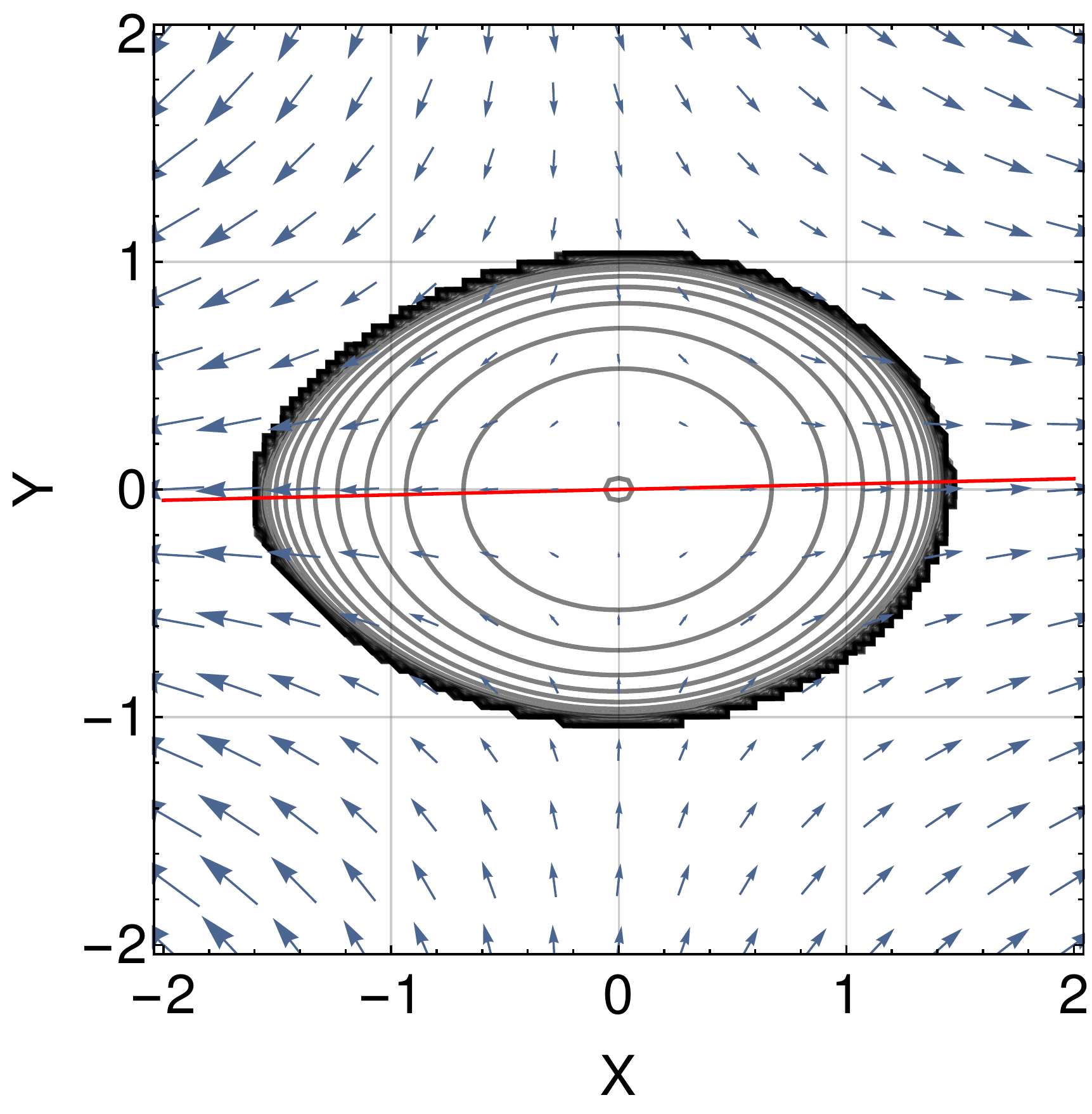}
		\caption{ $ \theta=1.5 $, $ n=1 $ }
		\label{fig.density-plots-n1-b}
	\end{subfigure}
	\begin{subfigure}{.32\textwidth}
		\includegraphics[width=\textwidth]{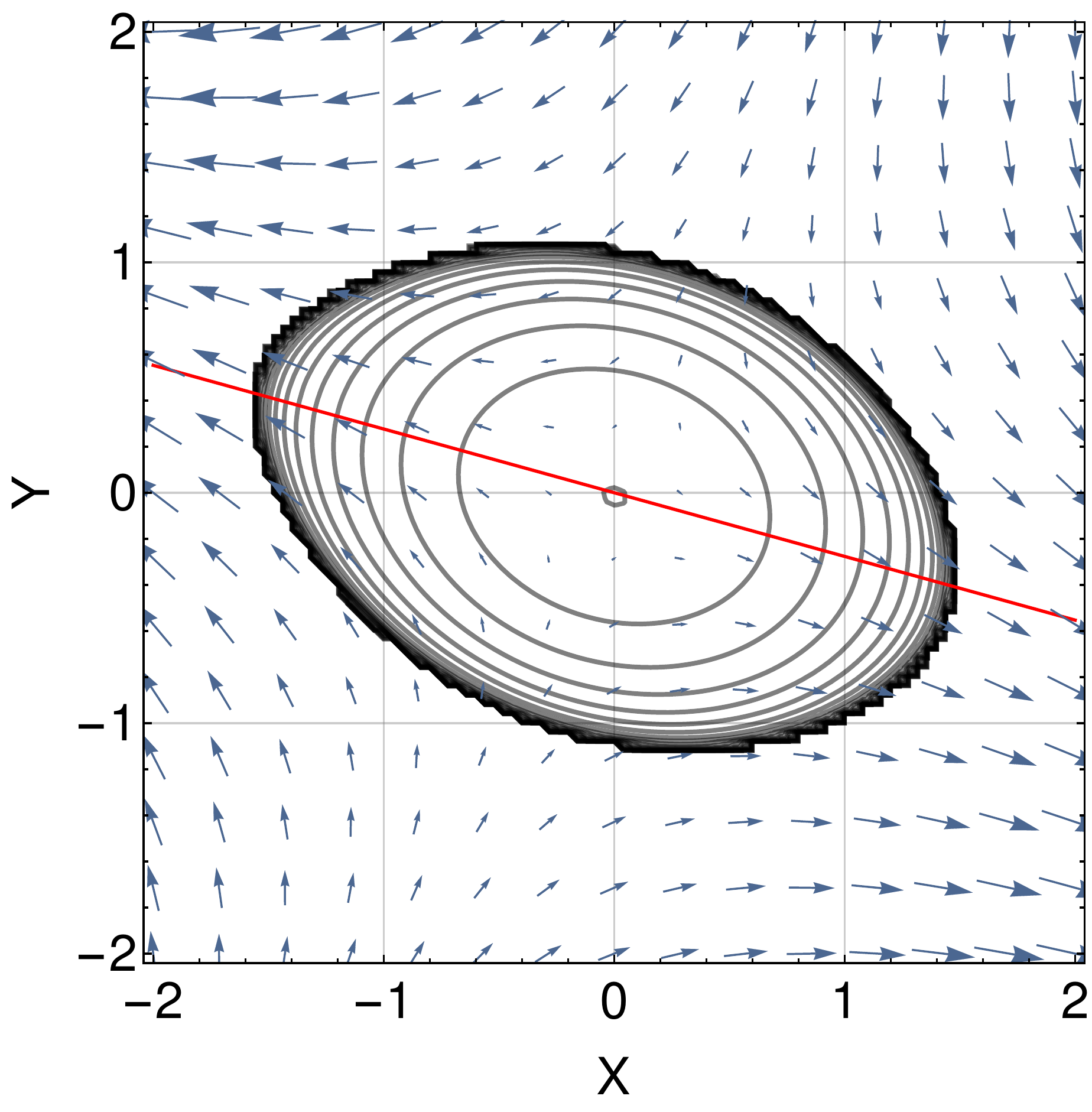}
		\caption{ $ \theta=2.5 $, $ n=1 $ }
		\label{fig.density-plots-n1-c}
	\end{subfigure}
	\caption{ Density contour plots of the star for $ M=1.0 $, $ a=0.9 $, $ r=6.0 $, $ L=1.0 $ and $ n=1 $. 
	The arrows indicate the resultant force field due to the tidal and the gravito-magnetic effects. The red lines show the radial $ r $-direction of the black hole.
	The constant density contour lines are obtained using the formula, $ \rho=\rho_c \times 10^{-0.2j} $, 
	where $ j=0,1,2,...,20 $. (a) represents a plot for $ \theta=0.5 $, (b) is plotted for $ \theta=1.5 $, and 
	(c) stands for $ \theta=2.5 $. Here, $ X $ denotes $ \tilde{x}^1 $, and $ Y $ represents $ \tilde{x}^2 $. }
	\label{fig.density-plots-n1}
	
\end{figure}

Now we will briefly comment on the density profile of the star near the tidal disruption.
Figure (\ref{fig.density-plots-n1}) shows the density contour plots of the star in $ \tilde{x}^1-\tilde{x}^2 $ plane 
plus the resultant force field (tidal and gravito-magnetic) at the critical limit of tidal disruption, i.e. at Roche limit. 
 The plots are obtained for parameter values $ M=1.0 $, $ a=0.9 $, $ r=6.0 $, $ L=1.0 $ and $ n=1 $. 
The density plots show that due to tidal effects a spherical star gets distorted making its structure 
asymmetric. As described in \cite{Ishii-Kerr}, on the equatorial plane, this asymmetry is introduced by the 
third and fourth order terms in tidal approximation which is what we also find.

\begin{figure}[H]
	\centering
	\begin{subfigure}{.32\textwidth}
		\includegraphics[width=\textwidth]{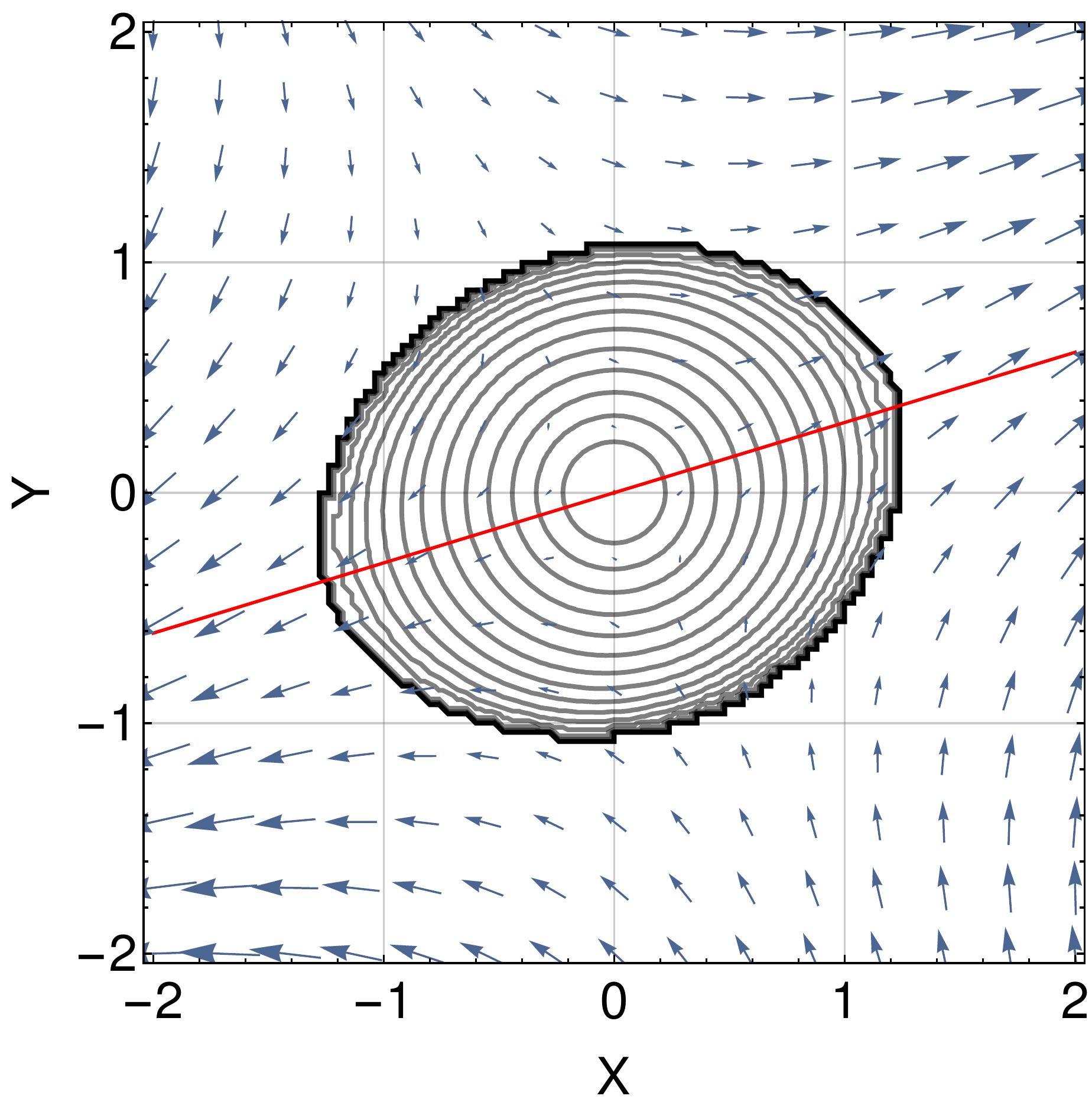}
		\caption{ $ \theta=0.5 $, $ n=3 $ }
		\label{fig.density-plots-n3-a}
	\end{subfigure}
	\begin{subfigure}{.32\textwidth}
		\includegraphics[width=\textwidth]{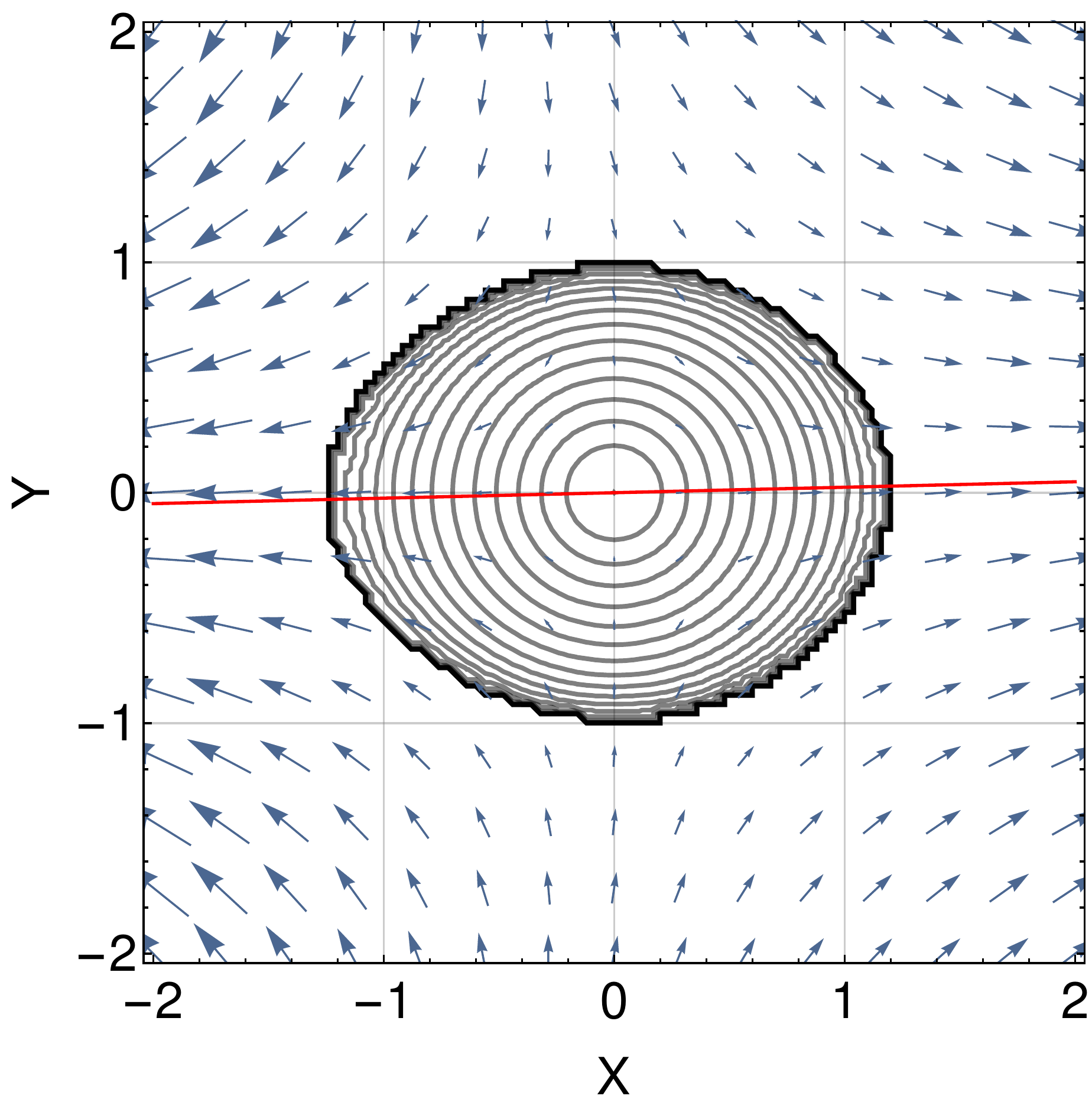}
		\caption{ $ \theta=1.5 $, $ n=3 $ }
		\label{fig.density-plots-n3-b}
	\end{subfigure}
	\begin{subfigure}{.32\textwidth}
		\includegraphics[width=\textwidth]{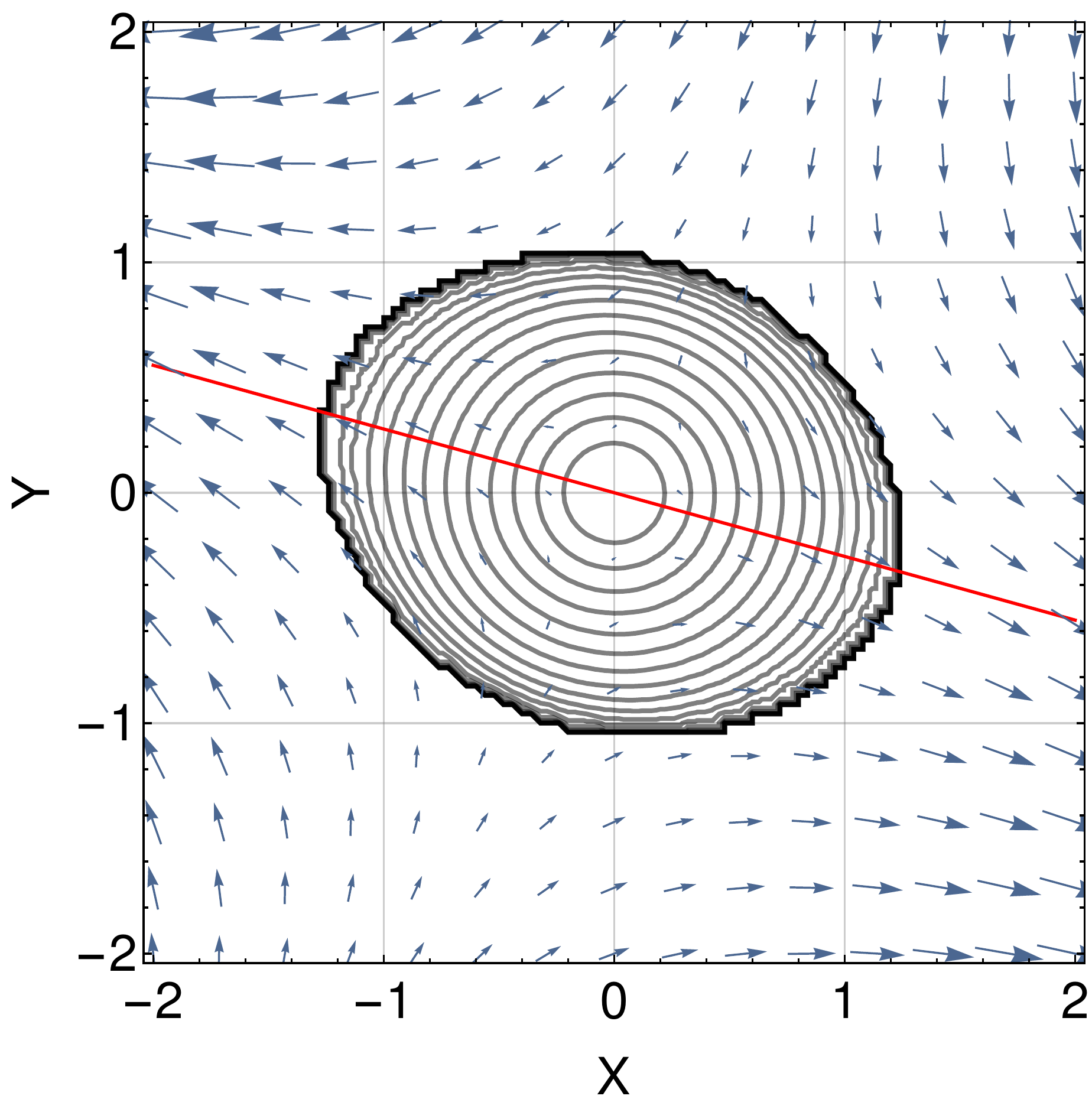}
		\caption{ $ \theta=2.5 $, $ n=3 $ }
		\label{fig.density-plots-n3-c}
	\end{subfigure}
	\caption{Density contour plots of the star for $ M=1.0 $, $ a=0.9 $, $ r=6.0 $, $ L=1.0 $ and $ n=3 $. 
	The arrows indicate the resultant force field due to the tidal and the gravito-magnetic effects. The red lines show the radial $ r $-direction of the black hole.
	The constant density contour lines, in this case, are obtained using the formula, 
	$ \rho=\rho_c \times 10^{-0.4j} $, where $ j=0,1,2,...,20 $. Similar to the previous case, (a) represents a 
	plot for $ \theta=0.5 $, (b) is plotted for $ \theta=1.5 $, and (c) stands for $ \theta=2.5 $. As before, 
	here $ X $ denotes $ \tilde{x}^1 $, and $ Y $ represents $ \tilde{x}^2 $. }
	\label{fig.density-plots-n3}
	
\end{figure}

An important feature of the tidal effects for non-equatorial orbits is its line of maximum deformation. 
This points downward in the upper hemisphere and upward in the lower hemisphere following the $ r $-direction, 
as seen from figs.(\ref{fig.density-plots-n1-a}, \ref{fig.density-plots-n1-c}) for $n=1$ and from figs.(\ref{fig.density-plots-n3-a}, 
\ref{fig.density-plots-n3-c}) for $n=3$. In these figures, the arrows indicate the magnitude and direction of the total force field (due to tidal and gravito-magnetic
effects). The star is deformed in accordance with the force field. Note that, the exact line of maximum deformation deviates from the $ r $-direction slightly for off-equatorial positions of the star. This might be due to the fact that the kerr geometry is not spherically symmetric.

The direction of this maximum deformation 
varies with the angular position ($ \theta $) of the star and the corresponding tilt in the density plots can be observed only if we look on 
the $ \tilde{x}^1-\tilde{x}^2 $ plane through the $ \tilde{x}^3 $-axis. Whereas, it is exactly aligned with the $ \tilde{x}^1 $-direction on the 
equatorial plane, i.e., for $ \theta=\pi/2 $.
In the $ \tilde{x}^1-\tilde{x}^3 $ plane, we will always see the star to be deformed along the $ \tilde{x}^1 $-direction only. 
We can also note that the deviation of the center of mass of the star from the origin is visible for $ n=1 $ and it is mostly along $ \tilde{x}^2 $ axis.

It is necessary to mention that for the co-rotational velocity field (eq.(\ref{eq.vel-field})) we have considered here, the gravito-magnetic force field can be greater in magnitude than that of tidal force field as we choose circular orbits with smaller $ r $ values. As a result the total force field can deform the star in such a way that the cusp forms on the other side of the star surface which is away from the black hole. 

In figure (\ref{fig.density-plots-n3}), we have shown the density contour plots (plus the resultant force field as before) 
for $ n=3 $ in the $ \tilde{x}^1-\tilde{x}^2 $ plane, taking the same values of other parameters as for the case $ n=1 $. 
The nature of the plots and the characteristic behavior of tidal force for non-equatorial orbits, as discussed in the previous 
case, are also found to be similar like the $ n=1 $ case. The only significant difference between the two cases is the fact 
that the amount of deformation of the stars is less for $ n=3 $ than that of $ n=1 $ for the same degree of tidal effects. 
Therefore, stars with higher equation of state parameter will have much stronger resistance against its deformation 
in shape before finally getting tidally disrupted completely.

\section{Summary and Discussion}
\label{sec-4}

In this paper, we have carried out an analysis of tidal effects on celestial objects in stable circular orbits away from the equatorial
plane, in Kerr black hole backgrounds. Our analysis is numerical, and involves incorporating constraints on such orbits in a Fermi 
normal coordinate system, where we have closely followed the related work reported in \cite{Ishii-Kerr} for equatorial circular orbits.

Our results show that there might be significant differences in the nature of tidal disruptions of stellar objects in circular orbits off
the equatorial plane, compared to those on it. In particular, we have seen that stars in pro-grade orbits
are more stable against tidal disruptions than their retro-grade counterparts, 
and that stability in off-equatorial orbits is maximum at the two extreme points 
of the orbit, and minimum at the equator. We have further seen that the numerical value of the 
tidal disruption limit depends strongly on the equation
of state for orbits away from the equatorial plane, and can vary by an order of magnitude, depending on the polytropic index. 
Finally, we have seen that the density contours are deformed along a direction that varies with the angular position of the star. The line of maximum deformation is slightly deviated from the $ r $-direction and is exactly aligned along $ r $ on the equatorial plane. As we have explained, this pheonomenon 
occurs due to the combination of forces from tidal and gravito-magnetic effects. 

It will be interesting to extend this analysis further to (slowly) rotating stars. It is well known that rotation introduces 
anisotropy in the stellar structure, and it will be interesting to see how such anisotropy is affected by tidal effects. Such 
an analysis might be substantially more complicated to perform compared to what has been reported here, but will 
nonetheless be an important issue for further research. 

\bibliographystyle{utphys}
\bibliography{References}

\providecommand{\href}[2]{#2}\begingroup\raggedright\begin{thebibliography}{10}

\bibitem{uv-flare1}
S.~Ayal, M.~Livio, and T.~Piran, ``Tidal disruption of a solar-type star by a
  supermassive black hole,'' \href{http://dx.doi.org/10.1086/317835}{{\em ApJ}
  {\bfseries 545} (Dec, 2000) 772--780},
  \href{http://arxiv.org/abs/astro-ph/0002499}{{\ttfamily
  arXiv:astro-ph/0002499}}.

\bibitem{uv-flare2}
T.~Bogdanović, M.~Eracleous, S.~Mahadevan, S.~Sigurdsson, and P.~Laguna,
  ``Tidal disruption of a star by a black hole: Observational signature,'' {\em
  The Astrophysical Journal} {\bfseries 610} no.~2, (2004) 707.
  \url{http://stacks.iop.org/0004-637X/610/i=2/a=707}.

\bibitem{GRB-BH-WD}
C.~L. Fryer, S.~E. Woosley, M.~Herant, and M.~B. Davies, ``{Merging white dwarf
  / black hole binaries and gamma-ray bursts},''
  \href{http://dx.doi.org/10.1086/307467}{{\em Astrophys. J.} {\bfseries 520}
  (1999) 650--660},
\href{http://arxiv.org/abs/astro-ph/9808094}{{\ttfamily arXiv:astro-ph/9808094
  [astro-ph]}}.

\bibitem{GRB-BH-NS}
H.~T. Janka, T.~Eberl, M.~Ruffert, and C.~L. Fryer, ``Black hole: Neutron star
  mergers as central engines of gamma-ray bursts,''
  \href{http://dx.doi.org/10.1086/312397}{{\em Astrophys. J.} {\bfseries 527}
  (1999) L39},
\href{http://arxiv.org/abs/astro-ph/9908290}{{\ttfamily arXiv:astro-ph/9908290
  [astro-ph]}}.

\bibitem{Rezzolla}
F.~Pannarale, A.~Tonita, and L.~Rezzolla, ``{Black hole-neutron star mergers
  and short GRBs: a relativistic toy model to estimate the mass of the
  torus},'' \href{http://dx.doi.org/10.1088/0004-637X/727/2/95}{{\em Astrophys.
  J.} {\bfseries 727} (2011) 95},
\href{http://arxiv.org/abs/1007.4160}{{\ttfamily arXiv:1007.4160
  [astro-ph.HE]}}.

\bibitem{tidal-literature1}
C.~R. Evans and C.~S. Kochanek, ``{The tidal disruption of a star by a massive
  black hole},'' \href{http://dx.doi.org/10.1086/185567}{{\em ApJL} {\bfseries
  346} (Nov, 1989) L13--L16}.

\bibitem{Marck-tidal-literature2}
J.~A. Marck, A.~Lioure, and S.~Bonazzola, ``Numerical study of the tidal
  interaction of a star and a massive black hole,'' {\em Astron. Astrophys.}
  {\bfseries 306} (Feb, 1996) 666,
\href{http://arxiv.org/abs/astro-ph/9505027}{{\ttfamily arXiv:astro-ph/9505027
  [astro-ph]}}.

\bibitem{tidal-literature3}
K.~Uryu and Y.~Eriguchi, ``Newtonian models for black hole-gaseous star close
  binary systems,''
  \href{http://dx.doi.org/10.1046/j.1365-8711.1999.02224.x}{{\em MNRAS}
  {\bfseries 303} (Feb, 1999) 329--342},
  \href{http://arxiv.org/abs/astro-ph/9808120}{{\ttfamily astro-ph/9808120}}.

\bibitem{Ishii-Numerical}
M.~Ishii and M.~Shibata, ``Numerical study of the tidal disruption of neutron
  stars moving around a black hole:- compressible jeans and roche problems -,''
  \href{http://dx.doi.org/10.1143/PTP.112.399}{{\em Progress of Theoretical
  Physics} {\bfseries 112} no.~3, (2004) 399--413}.

\bibitem{Shibata-tidal}
M.~Shibata, ``Relativistic roche-riemann problems around a black hole,''
  \href{http://dx.doi.org/10.1143/PTP.96.917}{{\em Progress of Theoretical
  Physics} {\bfseries 96} no.~5, (1996) 917--932}.

\bibitem{Marck-tidal}
J.-A. Marck, ``Solution to the equations of parallel transport in kerr
  geometry; tidal tensor,''
  \href{http://dx.doi.org/10.1098/rspa.1983.0021}{{\em Proceedings of the Royal
  Society of London A:} {\bfseries 385} no.~1789, (1983) 431--438}.

\bibitem{Fishbone}
L.~G. Fishbone, ``The relativistic roche problem. i. equilibrium theory for a
  body in equatorial, circular orbit around a kerr black hole,''
  \href{http://dx.doi.org/10.1086/152395}{{\em ApJ} {\bfseries 185} (Oct, 1973)
  43--68}.

\bibitem{Ishii-Kerr}
M.~Ishii, M.~Shibata, and Y.~Mino, ``{Black hole tidal problem in the Fermi
  normal coordinates},''
  \href{http://dx.doi.org/10.1103/PhysRevD.71.044017}{{\em Phys. Rev.}
  {\bfseries D71} (2005) 044017},
\href{http://arxiv.org/abs/gr-qc/0501084}{{\ttfamily arXiv:gr-qc/0501084
  [gr-qc]}}.

\bibitem{Manasse-Misner}
F.~K. Manasse and C.~W. Misner, ``Fermi normal coordinates and some basic
  concepts in differential geometry,''
\href{http://dx.doi.org/10.1063/1.1724316}{{\em J. Math. Phys.} {\bfseries 4}
  (1963) 735--745}.

\bibitem{Wilkins}
D.~C. Wilkins, ``{Bound Geodesics in the Kerr Metric},''
\href{http://dx.doi.org/10.1103/PhysRevD.5.814}{{\em Phys. Rev.} {\bfseries D5}
  (1972) 814--822}.

\bibitem{Hughes1}
S.~A. Hughes, ``Evolution of circular, nonequatorial orbits of kerr black holes
  due to gravitational-wave emission,''
  \href{http://dx.doi.org/10.1103/PhysRevD.61.084004}{{\em Phys. Rev. D}
  {\bfseries 61} (Mar, 2000) 084004},
\href{http://arxiv.org/abs/gr-qc/9910091}{{\ttfamily arXiv:gr-qc/9910091
  [gr-qc]}}.

\bibitem{Hughes2}
S.~A. Hughes, ``Nearly horizon skimming orbits of kerr black holes,''
  \href{http://dx.doi.org/10.1103/PhysRevD.63.064016}{{\em Phys. Rev. D}
  {\bfseries 63} (Feb, 2001) 064016},
\href{http://arxiv.org/abs/gr-qc/0101023}{{\ttfamily arXiv:gr-qc/0101023
  [gr-qc]}}.

\bibitem{Ryan}
F.~D. Ryan, ``{Effect of gravitational radiation reaction on nonequatorial
  orbits around a Kerr black hole},''
  \href{http://dx.doi.org/10.1103/PhysRevD.53.3064}{{\em Phys. Rev.} {\bfseries
  D53} (1996) 3064--3069},
\href{http://arxiv.org/abs/gr-qc/9511062}{{\ttfamily arXiv:gr-qc/9511062
  [gr-qc]}}.

\bibitem{Bardeen}
J.~M. Bardeen, W.~H. Press, and S.~A. Teukolsky, ``Rotating black holes:
  Locally nonrotating frames, energy extraction, and scalar synchroton
  radiation,'' \href{http://dx.doi.org/10.1086/151796}{{\em ApJ} {\bfseries
  178} (1972) 347--370}.

\bibitem{Hughes3}
E.~Barausse, S.~A. Hughes, and L.~Rezzolla, ``{Circular and non-circular nearly
  horizon-skimming orbits in Kerr spacetimes},''
  \href{http://dx.doi.org/10.1103/PhysRevD.76.044007}{{\em Phys. Rev.}
  {\bfseries D76} (2007) 044007},
\href{http://arxiv.org/abs/0704.0138}{{\ttfamily arXiv:0704.0138 [gr-qc]}}.

\bibitem{gravito-magnetic}
K.~S. Thorne, R.~H. Price, and D.~A. MacDonald, {\em The Membrane Paradigm}.
\newblock (Yale University, New Haven, CT, 1986).

\bibitem{Karmakar-Tapo}
T.~Karmakar and T.~Sarkar, ``Distinguishing between kerr and rotating jnw
  space-times via frame dragging and tidal effects,''
  \href{http://dx.doi.org/10.1007/s10714-018-2408-y}{{\em General Relativity
  and Gravitation} {\bfseries 50} no.~7, (Jun, 2018) 85}.

\end{thebibliography}\endgroup

\end{document}